\documentclass[a4paper,11pt]{article}
\usepackage{jheppub} 
\usepackage[T1]{fontenc} 
\RequirePackage{graphicx}
\usepackage{enumerate}
\usepackage{bm}
\usepackage{braket}
\usepackage{slashed}
\usepackage{comment}
\usepackage{color}

\title{The Exact WKB analysis for asymmetric scalar preheating}
\author[a]{Seishi Enomoto}
\affiliation[a]{School of Physics, Sun Yat-sen University, Guangzhou 510275, China}
\author[b]{Tomohiro Matsuda}
\affiliation[b]{Laboratory of Physics, Saitama Institute of Technology,
Fukaya, Saitama 369-0293, Japan}
\emailAdd{seishi@mail.sysu.edu.cn}
\emailAdd{matsuda@sit.ac.jp}

\abstract{Using the exact WKB analysis of the higher-order differential
equations, we analyze the asymmetry in dynamical particle
production of a complex scalar field. 
The solution requires the Stokes phenomena of the fourth-order
 differential equation. 
We found that the interference of different types of the 
Stokes phenomena causes matter-antimatter asymmetry.
We also showed a specific example where asymmetry is
forbidden in the exact calculation, but a false asymmetry 
appears in the perturbative expansion. }

\begin{document}

\maketitle
\section{Introduction}
One of the most important physical parameters of the Universe is the
observed ratio of the baryon number density to photon density.
The Planck result\cite{Planck:2018vyg} suggests that the ratio is given
by\cite{ParticleDataGroup:2020ssz}
\begin{eqnarray}
n_B/n_\gamma&=&(6.12\pm0.04)\times 10^{-10}.
\end{eqnarray}
The theoretical and phenomenological 
approach to explaining the ratio is called baryogenesis, which is
usually based on Sakharov's three conditions\cite{Sakharov:1967dj}:
(1) non-conservation of baryon number, (2) breaking of C and CP
invariance, (3) deviation from thermal equilibrium.
However, some interesting models of baryogenesis do not fit into
Sakharov's scenario.
The model, which is called ``spontaneous baryogenesis'', proposed the
essential idea of such scenarios\cite{Cohen:1987vi, Cohen:1988kt, Cohen:1991iu}.
The term ``spontaneous'' means
spontaneous breaking of underlying symmetry (e.g, the global $U(1)$
symmetry of the conservation of the baryon number), which leads to
\begin{eqnarray}
\label{eq-sb0}
{\cal L}_{sb}&\sim &(\partial_{\mu}\theta)J^\mu_B,
\end{eqnarray}
where $\theta$ is the phase of a field, and
the phase is related to the (pseudo)
Nambu-Goldstone field of the symmetry.
$J^\mu_B$ is the corresponding current.
The point of their observation is that the above interaction can lead to
nonconservation of the baryonic current of matter, but
the nonconservation disappears when $\theta$ is neither inhomogeneous nor
time-dependent.
Normally, the term in Eq.(\ref{eq-sb0}) is identified
with the chemical potential in the Hamiltonian density by using the relation
\begin{eqnarray}
\label{eq-idensb}
{\cal H}_{chem}=-{\cal L}_{sb},
\end{eqnarray}
but this is not an obvious relation, as is discussed in
Ref.\cite{Arbuzova:2016qfh}.
We will see shortly the point of Ref.\cite{Arbuzova:2016qfh} in Sec.\ref{subsec-chem}.

The idea of spontaneous baryogenesis can be found in the Affleck-Dine
baryogenesis\cite{Affleck:1984fy}, which is one of the most popular
scenarios of baryogenesis.
The Affleck-Dine baryogenesis considers rotational motion of a field, 
which implies generation of the baryon number.
Because of spatial instabilities of the Affleck-Dine field,
 the produced baryon number ($Q$) is believed to form
Q-balls\cite{Kasuya:1999wu,Enqvist:1998en}.

An important aspect of our analysis is that these models consider a
dynamical (time-dependent) scalar field with rotational motion, whereas
the dynamical production of particles due to such motion has not
been discussed using the Stokes phenomena.
The term "dynamical particle production" here refers to particle
generation from a vacuum, which is described using the Bogoliubov
coefficients and the Stokes phenomena (mixing of solutions of a
differential equation). 
Let us now look at the details of this problem by reviewing previous
works. 
Since the emphasis in this paper is on identifying the origin of
the asymmetry, the discussion is entirely qualitative.

\subsection{The role of the chemical potential in dynamical particle
  production}
\label{subsec-chem}
The spontaneous baryogenesis scenario considers Eq.(\ref{eq-sb0}) to
introduce the matter-antimatter asymmetry, but the validity of the
identification (\ref{eq-idensb}) is questionable\cite{Arbuzova:2016qfh}.
The point of the argument is that the transformation with respect to
the ``dynamical'' fields $\phi_a$ and their canonical momentum conjugate 
$\pi_a={\partial \cal L}/{\partial \dot{\phi}_a}$ gives
\begin{eqnarray}
{\cal H}=\sum_a \pi_a \dot{\phi}_a-{\cal L}.
\end{eqnarray}
Applying the rule to the ``field'' $\theta$ with ${\cal L}_{sb}$,
one cannot find the ``chemical potential'' because of $\pi_\theta$.
As is described in Ref.\cite{Arbuzova:2016qfh}, it could be possible to
find a contribution similar to the chemical potential, but the process
is far from trivial.

On the other hand, assuming that the field in motion is an external
field, for which the momentum conjugate $\pi_\theta$ can be neglected, one
will immediately find the desired ``chemical potential'' in the
Hamiltonian.

Since the above situation is almost the same for
fermions\cite{Arbuzova:2016qfh}, it is natural to think that the
``chemical potential'' used in the spontaneous baryogenesis scenario 
requires further study.
For us, this point is one of the primary reasons for considering
original equations of motion for particle production instead of using
an effective theory.

Let us see the meaning of the ``chemical potential'' of the spontaneous
baryogenesis scenario in more detail.
Suppose that the field in motion can be regarded as an external field.
In that case, we know\cite{Arbuzova:2016qfh} that ${\cal L}_{SB}$ can
introduce the 
conventional chemical potential in the Hamiltonian formalism.
Then, a naive question arises.
``Does this chemical potential bias dynamical particle production ?''
The answer to this naive question is far from trivial.
To find an answer  (not a no-go theorem) to this question, consider a simple solution of a
bosonic field when ${\cal L}_{SB}$ has a constant and homogeneous $\mu\equiv
\dot{\theta}$.
In contrast to naive speculation, this model does not allow chemical
potentials to bias particle production.

Consider the simplest scenario of bosonic preheating given by
the action\cite{Enomoto:2017rvc}
\begin{eqnarray}
S_0&=&\int d^4 x\sqrt{-g}\left[\partial_\mu\phi^*\partial^{\mu}\phi
-m^2 |\phi|^2+\xi R|\phi|^2 \right].
\end{eqnarray}
Using conformal time $\eta$, one can write the metric
$g_{\mu\nu}=a^2(\eta){\rm diag}(1,-1,-1,-1)$ and $R=-6\ddot{a}/a^3$,
where $a$ is the cosmological scale factor and the dot denotes
time-derivative with respect to the conformal time.
Defining a new field $\chi\equiv a\phi$, one will find a simple form
\begin{eqnarray}
S_0&=&\int d^4 x \left[|\dot{\chi}|^2
-\omega^2 |\chi|^2\right],
\end{eqnarray}
where
\begin{eqnarray}
\omega^2&\equiv& a^2m^2 +
\left(-\Delta + \frac{\ddot{a}}{a}(6\xi-1)\right).
\end{eqnarray}
Here $\Delta$ is the Laplacian.
Annihilation ($a,b$) and creation ($a^\dagger,b^\dagger$) operators of
``particle'' and ``antiparticle'' appear in the standard
decomposition\footnote{While this decomposition is correct for this
model, it needs to be modified for more general CP-breaking
interactions. We will see later that modifying this decomposition will
be the main point of discussion in this paper.}
\begin{eqnarray}
\label{eq-decom0}
\chi&=& \int \frac{d^3 k}{(2\pi)^{3/2}}\left[
h(\eta) a(\bm{k}) e^{i \bm{k}\cdot \bm{x}}
+g^*(\eta) b^\dagger(\bm{k}) e^{-i \bm{k}\cdot \bm{x}}\right].\nonumber\\
\end{eqnarray}
For our calculation, we introduce conjugate momenta
$\Pi^\dagger\equiv\dot{\chi}$,
which can be decomposed as
\begin{eqnarray}
\Pi^\dagger&=& \int \frac{d^3 k}{(2\pi)^{3/2}}\left[
\tilde{h}(\eta) a(\bm{k}) e^{i \bm{k}\cdot \bm{x}}
+\tilde{g}^*(\eta) b^\dagger(\bm{k}) e^{-i \bm{k}\cdot \bm{x}}\right].\nonumber\\
\end{eqnarray}
Following Ref.\cite{ZS-original}, we expand
$h, \tilde{h}$ (particles) and $g, \tilde{g}$ (antiparticles)
as
\begin{eqnarray}
h&=&\frac{e^{-i\int^\eta \omega d\eta'}}{\sqrt{2\omega}}A_h
+\frac{e^{i\int^\eta \omega d\eta'}}{\sqrt{2\omega}}B_h,\nonumber\\
\tilde{h}&=&\frac{-i\omega e^{-i\int^\eta \omega d\eta'}}{\sqrt{2\omega}}A_h
+\frac{i\omega e^{i\int^\eta \omega d\eta'}}{\sqrt{2\omega}}B_h,
\end{eqnarray}
and
\begin{eqnarray}
g&=&\frac{e^{-i\int^\eta \omega d\eta'}}{\sqrt{2\omega}}A_g
+\frac{e^{i\int^\eta \omega d\eta'}}{\sqrt{2\omega}}B_g,\nonumber\\
\tilde{g}&=&\frac{-i\omega e^{-i\int^\eta \omega d\eta'}}{\sqrt{2\omega}}A_g
+\frac{i\omega e^{i\int^\eta \omega d\eta'}}{\sqrt{2\omega}}B_g,
\end{eqnarray}
where $A_{h,g}$ and $B_{h,g}$ are known as the Bogoliubov coefficients.
(To avoid confusion, we are following the notations used in
Ref.\cite{Enomoto:2017rvc}, which are slightly different from the ones
used in this paper.)  
For further simplification, we introduce $\alpha_{h,g}$ and $\beta_h,g$, which are
defined as
\begin{eqnarray}
\alpha_{h,g}&\equiv& e^{-i\int^\eta\omega d\eta'}A_{h,g}\\
\beta_{h,g}&\equiv& e^{i\int^\eta\omega d\eta'}B_{h,g}.
\end{eqnarray}
Now the equation of motion can be written as
\begin{eqnarray}
\dot{h}-\tilde{h}&=&0\\
\dot{\tilde{h}}+\omega^2 h&=&0,
\end{eqnarray}
which are solved for $\dot{\alpha}$ and $\dot{\beta}$ as 
\begin{eqnarray}
\dot{\alpha}_h&=&-i\omega \alpha_h
 +\frac{\dot{\omega}}{2\omega}\beta_h\nonumber\\
\dot{\beta}_h&=&i\omega \beta_h
 +\frac{\dot{\omega}}{2\omega}\alpha_h.
\end{eqnarray}
The same calculation gives $\dot{\alpha}_g$ and $\dot{\beta}_g$.

Let us see what happens when a constant chemical potential is introduced.
After adding a chemical potential
\begin{eqnarray}
{\cal L}&=&\dot{\chi}\dot{\chi}^* -\omega^2 |\chi|^2
-i\mu_\chi \left(\chi \dot{\chi}^*-\chi^* \dot{\chi}\right),
\end{eqnarray}
we find
\begin{eqnarray}
\label{eq-of-mo-boson}
\ddot{\chi}-2i\mu_\chi\dot{\chi}+(\omega^2-i\dot{\mu}_\chi)\chi&=&0.
\end{eqnarray}
Two terms that might cause differences between $h$ and $g$ are
$-2i\mu_\chi\dot{\chi}$ and $-i\dot{\mu}_\chi \chi$.
If one assumes constant chemical potential, only the first term will
remain. 

As we have noted above, one can verify that the constant
chemical potential does not generate asymmetry in this model.
Here, the equation of motion becomes
\begin{eqnarray}
\dot{h}-\tilde{h}-i\mu_\chi h&=&0\nonumber\\
\dot{\tilde{h}}+\omega^2h -i\mu_\chi\tilde{h}&=&0.
\end{eqnarray}
One can solve these equations for $\dot{\alpha}$ and $\dot{\beta}$ to
find
\begin{eqnarray}
\dot{\alpha}_h&=&-i(\omega-\mu_\chi)\alpha_h
 +\frac{\dot{\omega}}{2\omega}\beta_h\nonumber\\
\dot{\beta}_h&=&\frac{\dot{\omega}}{2\omega}\alpha_h+i(\omega+\mu_\chi)\beta_h.
\end{eqnarray}
The same calculation for $g$ gives
\begin{eqnarray}
\dot{\alpha}_g&=&-i(\omega+\mu_\chi)\alpha_g
 +\frac{\dot{\omega}}{2\omega}\beta_g\nonumber\\
\dot{\beta}_g&=&\frac{\dot{\omega}}{2\omega}\alpha_g+i(\omega-\mu_\chi)\beta_g.
\end{eqnarray}
One might be tempted to speculate that the shift of $\omega\pm \mu_\chi$ 
will bias the particle production, but this speculation fails in the present model.
Using a simple calculation of $\frac{d|\beta_h|^2}{dt}$ and $\frac{d|\beta_h|^2}{dt}$,
one can verify that the evolution of $|\beta_h|^2$ and $|\beta_g|^2$ are
identical in this case, resulting in no asymmetry
production.
See ref.\cite{Enomoto:2017rvc} for more details.

We have seen that the chemical potential may not bias particle
production even if the field in motion is regarded as an external field.
The above calculation also indicates that $\dot{\mu}\ne 0$ may resolve
the degeneracy.
See also
Refs\cite{Kusenko:2014uta, Pearce:2015nga, Adshead:2015jza,
Adshead:2015kza, Enomoto:2021hfv}
for recent discussions on this topic.
In fact, when considering chemical potentials in cosmology, they should
be time-dependent, since $\mu$ is usually defined using a time-dependent
parameter.
Of course, if we consider the chemical potential of the external field
in a system of Boltzmann equations, the complications described above
for dynamical particle production do not appear; see also
Refs.\cite{Kusenko:2014lra,Yang:2015ida,Wu:2019ohx} for 
a recent discussion on the Higgs relaxation.

The rotational motion of a scalar field can be seen not only at
the preheating stage of natural
inflation\cite{Freese:1990rb,Adams:1992bn}, but also inside
Q-balls\cite{Coleman:1985ki,Kusenko:1997si} or other cosmological defects in
 motion\cite{Liu:1992tn,Matsuda:2002br,Matsuda:2001yi}.\footnote{Because of the 
 tuning of the vacuum energy in supergravity, one can expect that
 supersymmetric domain walls can naturally
 decay safely\cite{Matsuda:1998ms}.} 
Particles generated by the oscillation will decay when it
obtains a large mass from the oscillating field, breaking the coherent
growth.
If the particles do not decay, there could be
 ``trapping''\cite{Kofman:2004yc, Enomoto:2013mla} of the oscillating
 field, which makes the global calculation chaotic. 
In this paper, we do not discuss the quantitative calculation but focus
on the local event and seek the origin of the asymmetry.

\subsection{Asymmetry in preheating scenarios}
\label{subsec-asymp}
When fundamental parameters such as mass and interaction coefficients
become time-dependent, particles can be produced from the vacuum, and
there are various reasons why fundamental parameters can change during
cosmological evolution. Among them, particle production by inflaton
oscillations is known to be very important for solving the problem of
reheating the universe after inflation.\cite{Traschen:1990sw,
Kofman:1997yn,Felder:1998vq,Kofman:2004yc}. 

A semiclassical calculation of fermionic particle production by a
Nambu-Goldstone boson is performed in Ref.\cite{Dolgov:1994zq}.
They started with a simple model for a complex scalar field $\Phi$ and
two fermions $Q$ and $L$:
\begin{eqnarray}
S&=&\int d^4 x\left[
g^{\mu\nu}\partial_\mu \Phi^* \partial_\nu\Phi -V(\Phi^*\Phi)\right.\nonumber\\
&&\left.+i \overline{Q}\gamma^\mu\partial_\mu Q
+i \overline{L}\gamma^\mu\partial_\mu L
+(g \Phi \overline{Q}L+h.c.)
\right].
\end{eqnarray}
This action is invariant under a $U(1)$ symmetry.
In Ref.\cite{Dolgov:1994zq}, they have chosen
\begin{eqnarray}
\Phi\rightarrow e^{i\alpha}\Phi,&
Q\rightarrow e^{i\alpha}Q,&
L\rightarrow L.
\end{eqnarray}
To introduce spontaneous breaking of the global symmetry, the
potential is as
\begin{eqnarray}
V(|\Phi|)&=&\lambda\left[\Phi^*\Phi-f^2/2\right]^2,
\end{eqnarray}
which gives 
\begin{eqnarray}
\langle \Phi \rangle&=& fe^{i\phi/f}/\sqrt{2}
\end{eqnarray}
Just for simplicity of notation, the dimensionless angular field
$\theta\equiv \phi/f$ has been introduced to obtain the effective
Lagrangian for $\theta$:
\begin{eqnarray}
{\cal L}&=&\frac{f^2}{2}\partial_\mu \theta \partial^\mu \theta
+i \overline{Q}\gamma^\mu\partial_\mu Q\nonumber\\
&&+i \overline{L}\gamma^\mu\partial_\mu L
+(g f\overline{Q}L e^{i\theta}+h.c.),
\end{eqnarray}
where the global symmetry is realized as
\begin{eqnarray}
\theta\rightarrow \theta+\alpha,&
Q\rightarrow e^{i\alpha}Q,&
L\rightarrow L.
\end{eqnarray}
For $\alpha=-\theta$, one can rewrite the Lagrangian as
\begin{eqnarray}
{\cal L}&=&\frac{f^2}{2}\partial_\mu \theta \partial^\mu \theta
+i \overline{Q}\gamma^\mu\partial_\mu Q
+i \overline{L}\gamma^\mu\partial_\mu L\nonumber\\
&&+(g f\overline{Q}L +h.c.) + \partial_\mu\theta J^\mu -U(\theta),
\end{eqnarray}
where $J^\mu=\overline{Q}\gamma^\mu Q$ denotes the fermion current of
the $U(1)$ symmetry, and $U(\theta)$ is assumed to be given by
\begin{eqnarray}
U(\theta)&=&\Lambda^4 \left[
1\pm \cos\theta\right].
\end{eqnarray}
Note that the model is inspired by the model of natural
inflation\cite{Freese:1990rb,Adams:1992bn} and aims to propose
baryogenesis caused by the preheating stage after natural inflation.

The baryon number asymmetry generated by the Pseudo-Nambu Goldstone
Boson(PNGB) was calculated in Ref.\cite{Cohen:1987vi}.
Their result is 
$|\dot{n}_B|=\Gamma f^2 |\dot{\theta}|$, where $\Gamma$ is the decay
rate of the PNGB.
However, the authors of
Ref.\cite{Dolgov:1994zq} alerted that the original calculation might be
naive.
Note that a similar question may arise for the decay of the Affleck-Dine
field and a Q-ball.
In Ref.\cite{Dolgov:1996qq}, they have expanded the calculation to find
that the baryon number should be given by
\begin{eqnarray}
n_B&\propto&\Gamma f^2 \theta_i^3,
\end{eqnarray}
where the $\theta$-dependence is modified.
Their calculation in Ref.\cite{Dolgov:1996qq} uses the Bogoliubov
transformation and two different kinds of expansion: small $g$ and
small $\theta(t)$.
We will see later that the expansion of Ref.\cite{Dolgov:1996qq} 
may destroy the structure of the Stokes lines, to change the qualitative
analysis of the model.
What is important in the approach\cite{Dolgov:1996qq} is that they have
explicitly included interaction between different species $Q$ and $L$.

Before going forward, we explain why the exact WKB (EWKB) is crucial for
dynamical particle production.
The equations of motion become higher-order differential equations when
an interaction is introduced. (See the model described above.)
Global asymptotic analysis of the higher-order differential equations was
thought to be impossible to construct before the discovery of the ``new
Stokes lines'' by H.L. Berk, W.M. Nevins, and K.V. Roberts\cite{NBR}.
Later by T. Aoki, T. Kawai, and Y. Takei\cite{Virtual:2015HKT}, the notion of a
virtual turning point was discovered by applying 
microlocal analysis to Borel transformed WKB
solutions.
Since the virtual turning point cannot be detected by ordinary WKB solutions,
the conversion of the study to the one in a different space, 
the Borel plane on which the Borel transformed WKB solutions are
analyzed, was indispensable.
We are using these ideas to discuss matter-antimatter asymmetry in a
scalar preheating scenario with complex time-dependent parameters.

We are aware that the EWKB analysis is not yet common; one may be puzzled by
the claim that the Borel resummation of the WKB expansion is not an
approximation but gives an exact result. 
One might also wonder why the Stokes lines in the EWKB analysis are giving
an exact result when they are computed from the first term of the
expansion. Also, Stokes phenomena in ordinary differential equations may
not be a popular topic in cosmology. 
Therefore, to avoid confusion, we will review these topics in this
paper. 
Our discussions in the review sections are based on our previous
papers\cite{Enomoto:2020xlf,Enomoto:2021hfv} and the
textbook\cite{Virtual:2015HKT}.
We also refer the reader to Ref.\cite{Taya:2020dco,Kitamoto:2020tjm,
Hashiba:2021npn}. 
On the other hand, we do not review the mathematical details of the
formulation.
For details and proofs, see Refs.\cite{Virtual:2015HKT, Voros:1983,
Delabaere:1993}.
We only describe how to use this useful tool in a cosmological particle
generation scenario.

\subsection{The Stokes phenomena and the Bogoliubov transformation in
  cosmological particle creation}
\label{subsec-invst}
First, we explain why the Bogoliubov transformation is explained by 
the Stokes phenomena, applying them to the standard preheating scenario.

The motion of the inflaton field is a damped oscillation.
However, at least near the center of the oscillation, where particle
production is likely to take place, a linear approximation with respect to
$t$ can be made.
Typically, the mass of a scalar field (e.g, $\chi$) is supposed to be
given by 
\begin{eqnarray}
m^2_\chi(t)&=&m_0^2 +g^2_2 \phi(t)^2,
\end{eqnarray}
where $\phi(t)$ is the oscillating inflaton field.
If we consider the Lagrangian given by
\begin{eqnarray}
{\cal L}_\chi&=&\frac{1}{2}\partial_\mu \chi\partial^\mu \chi
 -\frac{1}{2}m_0^2\chi^2 -\frac{1}{2}g_2^2 \phi(t)^2\chi^2,
\end{eqnarray}
the equation of motion becomes
\begin{eqnarray}
\frac{d^2 \chi}{dt^2}+\left[k^2+m^2_\chi(t)\right]\chi=0.
\end{eqnarray}
If one replaces $\phi(t)$ with $\phi(t)\simeq vt$, the above equation is
equivalent to the Schr\"odinger equation of the scattering problem.
Following the mathematician's formulation, in which ``$m$'' in the
original Schr\"odinger equation is removed by setting ``$2m=1$'', we define
inverted quadratic ``potential'' given by
\begin{eqnarray}
V(t)&=&-\left(g_2^2v^2\right)t^2,
\end{eqnarray}
where the corresponding ``energy'' is 
\begin{eqnarray}
E&=&k^2+m_0^2.
\end{eqnarray}
Note that $E>V$ is always true in this case.
Therefore, there is no classical turning point in the scattering
problem.

There are a variety of methods for finding the wave functions of the
general one-dimensional scattering problem of quantum mechanics.
For the inverted quadratic potential, one can find the exact solution 
(i.e, the Weber function, or the parabolic cylinder functions).
Analytic continuation of the WKB expansion (complex WKB) and related
topics have a long history\cite{Pokrovskii:1961}.
For instance, it has been applied to pair production in a vacuum by an alternating
field\cite{Brezin:1970xf}.
Note however that without the Borel resummation the definitions of the WKB
expansion and the analytic continuation are vague.
The most obvious confusion will be that without the Bore resummation the
Stokes lines seem to be changed by introducing higher terms.
Therefore, the EWKB is crucial for our argument.

Since the model gives the typical structure of the Stokes lines
(the so-called ``Merged pair of simple Turning Points(MTP)''),
we are going to show explicitly the calculation in detail.

Typically, the (conventional) WKB expansion is used to
find
\begin{eqnarray}
\label{eq-WKB}
\chi_k(t)&=& \frac{\alpha_k(t)}{\sqrt{2\omega_k}}e^{-i \int^t \omega dt}
+\frac{\beta_k(t)}{\sqrt{2\omega}}e^{+i \int^t \omega dt},
\end{eqnarray}
where
\begin{eqnarray}
\omega^2_k(t)&\equiv&k^2+m_\chi^2(t).
\end{eqnarray}
Here $\alpha_k=1$ and $\beta_k=0$ are considered for the initial vacuum state.
These solutions are the asymptotic states of the exact solutions 
for which the number densities are defined. 
(Of course, in cosmology one is not considering exact $t\rightarrow \pm
\infty$ limits for the states.
The Stokes phenomenon is a local event.)
The distribution of the particle in the final state is 
\begin{eqnarray}
n_\chi(k)=|\beta_k|^2,
\end{eqnarray}
which can be found by solving the scattering problem of the
corresponding Schr\"odinger equation.
The connection formulae of the solutions, which give $\beta_k\ne 0$ for
the final state, are nothing but the consequence of the Stokes phenomena.

To see what happens, we are going to solve the
equation of motion explicitly using the Weber function.
For the above model (i.e, scattering by the inverted quadratic
potential), the following Weber equation
\begin{eqnarray}
y''(z)+\left(\nu+\frac{1}{2}-\frac{1}{4}z^2\right)y(z)=0
\end{eqnarray}
has the solution $D_\nu(z), D_{-\nu-1}(iz)$.\footnote{Note that the
following relation 
\begin{eqnarray}
D_\nu(z)&=&e^{i\nu\pi}D_\nu(-z)+\frac{\sqrt{2\pi}}{\Gamma(-\nu)}
 e^{i(\nu+1)\pi/2}D_{-\nu-1}(-iz)\nonumber
\end{eqnarray}
shows that both $D_\nu(-z)$ and $D_{-\nu-1}(iz)$ are also the solutions
of the equation.}
More specifically, one can define
\begin{eqnarray}
z&\equiv& ie^{i\pi/4}\sqrt{2g_2v}t
\end{eqnarray}
in the original field equation to find 
\begin{eqnarray}
\frac{d^2 \chi}{dz^2}+\left[\nu+\frac{1}{2} -\frac{1}{4}z^2\right]\chi=0.
\end{eqnarray}
Here we defined
\begin{eqnarray}
\nu=\frac{k^2+m_0^2}{2g_2v}i-\frac{1}{2},
\end{eqnarray}
and for later use we define 
\begin{eqnarray}
\kappa&\equiv& \frac{k^2+m_0^2}{2g_2v}
\end{eqnarray}
and 
\begin{eqnarray}
\nu=i\kappa -\frac{1}{2}.
\end{eqnarray}
Here, $\kappa$ is an important parameter, which is later used to
 estimate particle production.
In the next section, we will show how to calculate $\kappa$ in the EWKB.
The asymptotic forms of the Weber function at $|z|\rightarrow \infty$ are given by
\begin{eqnarray}
1.&|\mathrm{arg} z|<\frac{3\pi}{4}, & \,\,D_\nu(z)\rightarrow e^{-z^2/4}z^\nu,\\
2.& -\frac{5}{4}\pi<\mathrm{arg} z<-\frac{\pi}{4},&\,\, D_\nu(z)\rightarrow
 e^{-\frac{z^2}{4}}z^\nu-\frac{\sqrt{2\pi}}{\Gamma(-\nu)}e^{-i\nu\pi+\frac{z^2}{4}}z^{-\nu-1}, \\
3. & \frac{\pi}{4}<\mathrm{arg} z<\frac{5\pi}{4},&\,\, D_\nu(z)\rightarrow
 e^{-\frac{z^2}{4}}z^\nu-\frac{\sqrt{2\pi}}{\Gamma(-\nu)}e^{i\nu\pi+\frac{z^2}{4}}z^{-\nu-1}.
\end{eqnarray}
Since $z\equiv i e^{i\pi/4}\sqrt{2g_2v}t$ is used here, $t<0$
gives $\frac{5\pi}{4}<\mathrm{arg}z<\frac{9\pi}{4}$, which corresponds
to region 1.
Also, $t\rightarrow +\infty$ corresponds to region 3.
Therefore, we find for $t\rightarrow -\infty$,
\begin{eqnarray}
e^{-\frac{z^2}{4}}&=&e^{-i\frac{g_2v}{2}t^2}\\
z^\nu&=&e^{(i\kappa-\frac{1}{2})\log z}\nonumber\\
&=&e^{(i\kappa-\frac{1}{2})\left(\log(\sqrt{2g_2 v} |t|)+i\frac{3\pi}{4}\right)},
\end{eqnarray}
which gives ($t=-|t|=e^{\pi i}|t|$ is used here)
\begin{eqnarray}
D_\nu(z)&\simeq&
e^{-i\frac{g_2v}{2}t^2}e^{(i\kappa-\frac{1}{2})\left(\log(\sqrt{2g_2 v}
|t|)-i\frac{\pi}{4}\right)},\\
D_{-\nu-1}(iz)&\simeq& e^{+i\frac{g_2v}{2}t^2}e^{(-i\kappa-\frac{1}{2})\left(\log(\sqrt{2g_2 v}
						 |t|)+i\frac{\pi}{4}\right)}.
\end{eqnarray}

Note that the above solutions in the limit $t\rightarrow -\infty$ are
giving the $\pm$ WKB solutions of Eq.(\ref{eq-WKB}).
Therefore, we define 
\begin{eqnarray}
\chi_- &\rightarrow&D_{\nu}(z)\\
\chi_+ &\rightarrow&D_{-\nu-1}(iz).
\end{eqnarray}
On the other hand, in the $t\rightarrow +\infty$ limit, we find
\begin{eqnarray}
e^{-\frac{z^2}{4}}&=&e^{-i\frac{g_2v}{2}t^2}\\
z^\nu&=&e^{(i\kappa-\frac{1}{2})\log z}\nonumber\\
&=&e^{(i\kappa-\frac{1}{2})\left(\log(\sqrt{2g_2 v} t)  +i\frac{3\pi}{4}\right)},
\end{eqnarray}
which gives in this limit,
\begin{eqnarray}
D_\nu(z)&\simeq& e^{-i\frac{g_2v}{2}t^2}e^{(i\kappa+\frac{1}{2})
\left(\log(\sqrt{2g_2 v}t)+i\frac{3\pi}{4}\right)}\nonumber\\
&&+i\frac{\sqrt{2\pi}}{\Gamma(-\nu)}e^{i\frac{g_2v}{2}t^2}
e^{-\kappa \pi}e^{(-i\kappa-\frac{1}{2})\left(\log(\sqrt{2g_2 v}t)
+i\frac{3\pi}{4}\right)}.\nonumber\\
\end{eqnarray}
Immediately, one will find that in the $t=+\infty$ limit the asymptotic
form of the exact solution $D_\nu(z)$ is the mixture of the
$\pm$ WKB solutions, which gives the connection formula of the Stokes phenomena.
In this case, the connection formula gives the Bogoliubov
transformation of the WKB solutions.
In the calculation of the connection formula, we use
\begin{eqnarray}
\Gamma(z)\Gamma(1-z)&=&\frac{\pi}{\sin \pi z}\\
\Gamma(\bar{z})&=&\overline{\Gamma(z)}\\
1+\nu&=&1+\left(i\kappa-\frac{1}{2}\right)=-\overline{\nu}
\end{eqnarray}
for the calculation of $\Gamma(-\nu)=\Gamma(-i\kappa+\frac{1}{2})$.
This gives 
\begin{eqnarray}
|\Gamma(-\nu)|^2&=&\frac{\pi}{\sin \pi (-\nu)}\nonumber\\
&=&\frac{2\pi i}{e^{-i\pi\nu}-e^{i\pi\nu}}\nonumber\\
&=&\frac{2\pi}{e^{\pi\kappa}+e^{-\pi\kappa}}\\
\Gamma(-\nu)&=&\frac{\sqrt{2\pi}e^{-\pi\kappa/2}}{\sqrt{1+e^{-2\pi\kappa}}}
e^{\mathrm{arg}\Gamma(-\nu)}.
\end{eqnarray}
Finally, one obtains the connection formula given by
\begin{eqnarray}
\left(
\begin{array}{c}
\alpha_k^R\\
\beta_k^R
\end{array}
\right)
&=&
\left(
\begin{array}{cc}
\sqrt{1+e^{-2\pi \kappa}}e^{i\theta_1} & ie^{-\pi\kappa+i\theta_2}\\
-ie^{-\pi\kappa-i\theta_2} &\sqrt{1+e^{-2\pi \kappa}}e^{-i\theta_1} 
\end{array}
\right)
\left(
\begin{array}{c}
\alpha_k^L\\
\beta_k^L
\end{array}
\right),\nonumber\\
\end{eqnarray}
where L and R denote $t\rightarrow -\infty$ and $t\rightarrow +\infty$, respectively.
Here, all phase parameters are put into $\theta_{1,2}(k)$.
Viewing the result as the solution of the scattering problem, the
reflection and the penetration amplitudes are given by
\begin{eqnarray}
|R_k|&=&\frac{e^{-\pi \kappa}}{\sqrt{1+e^{-2\pi \kappa}}}\nonumber\\
|T_k|&=&\frac{1}{\sqrt{1+e^{-2\pi \kappa}}}.
\end{eqnarray}

In the above scenario, there is no classical reflection point.
This means that classically reflection is not allowed in the
scattering problem. 
Therefore, particle production becomes significant when $\kappa \ll1$,
where the quantum scattering process becomes significant.

\subsection{Introduction to the Exact WKB analysis}
The Wentzel - Kramers - Brillouin - Jeffreys(WKB or WKBJ) approximation
is a well-known method for solving linear differential equations.
Although the first term of the expansion normally gives an
excellent approximate solution, the method is formally giving a
divergent power series.
The Exact WKB analysis considers the Borel resummation to solve this problem.
Thanks to the Borel resummation, the Borel sum is extended to the
complex $\eta$-plane, where $\eta$ is the expansion parameter
corresponding to $\hbar^{-1}$.
Using these ideas of the EWKB, one can guarantee calculations of the WKB
expansion avoiding the infamous problems of the original formulation. 

The typical mathematician's formulation of the EWKB uses $\eta\equiv
 \hbar^{-1}\gg 1$ instead of $\hbar$.
Following Ref.\cite{Virtual:2015HKT}, our starting point is the
mathematician's ``Schr\"odinger equation'' given by 
\begin{eqnarray}
\left[-\frac{d^2}{dx^2}+\eta^2 Q(x)
\right]\psi(x,\eta)&=&0.
\end{eqnarray}
Introducing the ``potential'' $V$ and the ``energy'' $E$, we define
\begin{eqnarray}
Q(x)&=&V(x)-E.
\end{eqnarray}
Writing the solution as $\psi(x,\eta)=e^{R(x,\eta)}$,
we have 
\begin{eqnarray}
\psi&=&e^{\int^x_{x_0}S(x,\eta)dx}
\end{eqnarray}
for $S(x,\eta)\equiv \partial R/\partial x$.
Here $x_0$ is an arbitrary parameter but normally taken on a turning point.
For $S$, we have 
\begin{eqnarray}
-\left(S^2 +\frac{\partial S}{\partial x}\right)+\eta^2 Q&=&0.
\end{eqnarray}
If one expands $S$ as $S(x,\eta)=\sum_{n=-1}^{n=\infty}\eta^{-n} S_{n}$,
one will find
\begin{eqnarray}
S=\eta S_{-1}(x)+ S_0(x)+\eta^{-1}S_1(x)+...,
\end{eqnarray}
which leads
\begin{eqnarray}
\label{eq_firstterm}
S_{-1}^2&=&Q\\
2S_{-1}S_j&=&-\left[\sum_{k+l=j-1,k\ge 0,l\ge 0}S_kS_l + \frac{d
	       S_{j-1}}{dx}\right]\\
&&(j\ge 0).\nonumber
\end{eqnarray}
Using the relation between the odd and the even series, one will have 
\begin{eqnarray}
\label{eq-sodd}
\psi&=&\frac{1}{\sqrt{S_{odd}}}e^{\int^x_{x_0}S_{odd}dx}\\
&&S_{odd}\equiv\sum_{j\ge 0}\eta^{1-2j}S_{2j-1}.
\end{eqnarray}
Depending on the sign of the first $S_{-1}=\pm \sqrt{Q(x)}$, there are
two solutions $\psi_\pm$, which are given by
\begin{eqnarray}
\psi_{\pm}&=&\frac{1}{\sqrt{S_{odd}}}\exp\left(\pm \int^x_{x_0}S_{odd}
					  dx\right)\nonumber\\
&=&e^{\pm \eta \int\sqrt{Q}dx}\sum^\infty_{n=0}\eta^{-n-1/2}\psi_{\pm,n}(x).
\end{eqnarray}
The above WKB expansion is usually divergent.
The Borel transform is defined by
\begin{eqnarray}
\psi_{\pm}^B&=&\sum^\infty_{n=0}
\frac{\psi_{\pm,n}(x)}{\Gamma(n+\frac{1}{2})}\left(y\pm 
 s(x)\right)^{n-\frac{1}{2}}.
\end{eqnarray}
Note that the shift of the integral of the inverse-Laplace integration
of the Borel sum is determined by
$S_{-1}$ as
\begin{eqnarray}
\label{eq-yspace}
\psi_\pm &\rightarrow&\Psi_\pm\equiv\int^\infty_{\mp
 s(x)}e^{-y\eta}\psi_\pm^B(x,y)dy,\\
\label{eq_sx}
&&s(x)\equiv \int^x_{x_0}S_{-1}(x)dx,
\end{eqnarray}
where the $y$-integral is parallel to the real axis.
Note also that the Borel transform ($\psi^B_\pm$) 
corresponds to the Laplace transformation 
with respect to the expansion parameter $\eta$.

At this stage, the reader may think that the Stokes line computed from
$s(x)$ is merely an approximation and not an exact result. 
We emphasize here that this speculation is incorrect and that the Stokes
lines in the EWKB are exact. 
One may also be puzzled by the Laplace transform (Borel
transform) for the expansion parameter $\eta\equiv \hbar^{-1}$. 
It is not trivial, but after
some mathematics, one will see that the Stokes lines are exact and that
continuation with respect to $\eta$ is possible with this
formalism\cite{Virtual:2015HKT}.

\begin{figure}[t]
\centering
\includegraphics[width=.45\textwidth]{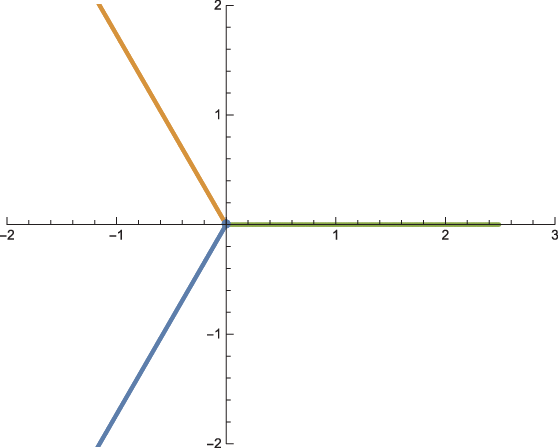}
 \caption{The Stokes lines of the Airy function, which are the solutions
 of $\mathrm{Im} [s(x)]=0$, are plotted 
 on the complex $x$-plane. 
The classical turning point of the Schr\"odinger equation
is shown at the origin.}
\label{fig-airy}
\end{figure}

Let us see how the Stokes phenomena work in this formalism.
For simplicity, we refer to the familiar Airy function ($Q(x)=x$) here.
On the complex $x$-plane, three Stokes lines are coming out of a turning
point, which appears at $x=0$ and corresponds to the classical turning
point of the Schr\"odinger equation.
We show the Stokes lines and the turning point in Fig.\ref{fig-airy}.
The Stokes lines are the solutions of $\mathrm{Im} [s(x)]=0$.

\begin{figure}[t]
\centering
\includegraphics[width=.45\textwidth]{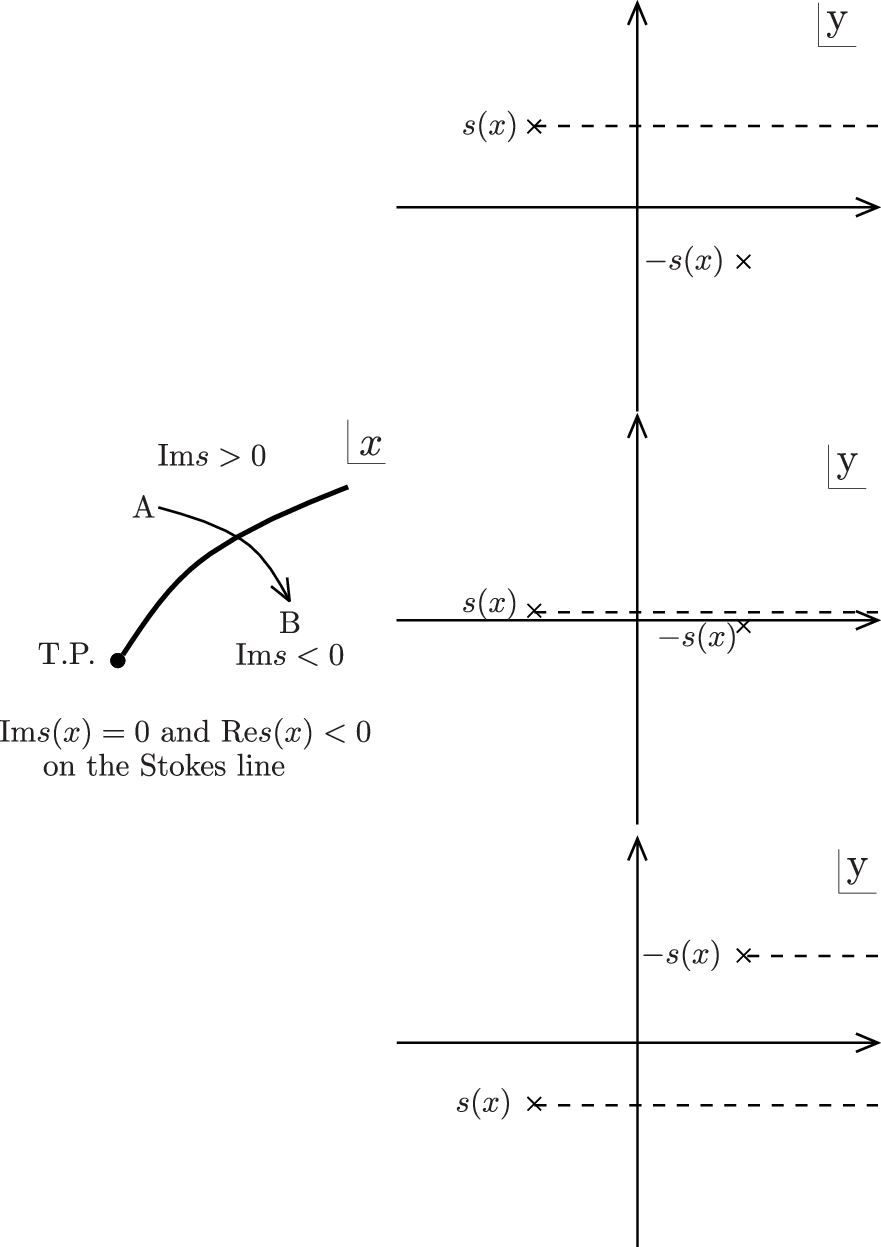}
 \caption{The Stokes phenomenon is explained by the $y$-integration.
On the left panel, a Stokes line (the other two are omitted) 
coming out from a turning point is shown.
The Stokes line is drawn on the complex $x$-plane.
We take Re$[s(x)]<0$ for the Stokes line.
On the right three panels, the contour of the $y$-integration in
 Eq.(\ref{eq-yspace}) is shown.
If the solution, whose integration starts from $s(x)$,
moves from $A$ to $B$ crossing the Stokes line, the
 integration contour finally picks up the integration starting from $-s(x)$.  }
\label{fig-EWKB}
\end{figure}

To understand the Stokes phenomena of the exact WKB analysis, consider
the $y$-integration in Eq.(\ref{eq-yspace}) for the $\pm$ solutions.
Plotting the integration paths on the $y$-plane,
one can see that the two paths of the $\pm$ solutions overlap when $x$
is on the Stokes line.
(Remember that $\mp s(x)$ is giving the starting point of the $y$-integration
and $\mathrm{Im} [s(x)]=0$ is the definition of the Stokes line.)
Therefore, one of these solutions (on the left)
 will develop additional contributions as $x$ goes across the Stokes
 line.
The situation is shown in Fig.\ref{fig-EWKB}.
In this case, the solution on the left picks up the other's
integration to give the Stokes phenomenon\cite{Virtual:2015HKT}.
Using the above idea, one can find the connection formulae given by
\begin{itemize}
\item Crossing the $\psi_+$ Dominant Stokes line
      with an anticlockwise rotation (seen from the turning
      point)
\begin{eqnarray}
\Psi_+&\rightarrow& \Psi_+ +i\Psi_-\\
\Psi_-&\rightarrow& \Psi_-
\end{eqnarray}
\item Crossing the $\psi_-$-Dominant Stokes line with an anticlockwise rotation (seen from the turning
      point)
\begin{eqnarray}
\Psi_-&\rightarrow& \Psi_- +i\Psi_+\\
\Psi_+&\rightarrow& \Psi_+
\end{eqnarray}
\item An inverse rotation gives a minus sign in front of $i$.
\end{itemize}
The ``dominant'' solution in Fig.\ref{fig-EWKB} is the one appearing on
the left, whose integration contour picks up additional integration
after the Stokes phenomenon. 

\begin{figure}[t]
\centering
\includegraphics[width=.55\textwidth]{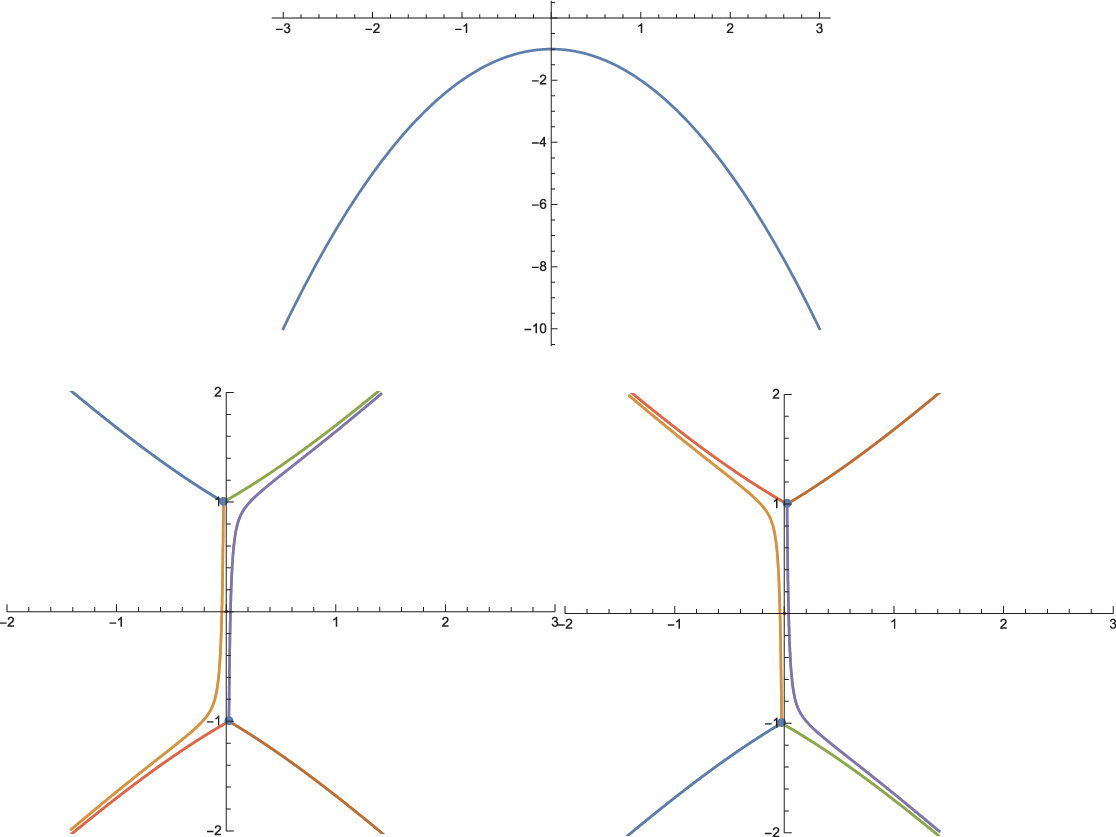}
 \caption{In the upper panel, the potential ($V(x)=-1-x^2$) is shown
 on the $V-x$ plane. 
In the lower panel, the Stokes lines of $Q(x)=-1-x^2\pm 0.05i$ are shown
on the complex $x$-planes.
One can see that there is a
 discontinuity between $Q(x)=-1-x^2 +0.05i$(lower left) and
 $Q(x)=-1-x^2- 0.05i$(lower right)}
\label{fig-invpot}
\end{figure}

As long as the Stokes lines are not degenerate and no singularities
appear, the above connection formulae are versatile. 
However, in some cases (e.g., scattering due to inverse quadratic potentials), the Stokes
lines become degenerate and the above formulae must be reconsidered. 
The degeneracy can be solved by introducing a small imaginary factor in $\eta$,
but the problem is that the formulation becomes discontinuous with
respect to the sign of the introduced imaginary factor. 
The situation is illustrated in Fig.\ref{fig-invpot}.
In this case, the normalization factor must be non-trivial to be
theoretically consistent with the $\pm$ sign of the imaginary part
of $\eta$.
The easiest way to find the normalization factor (this factor is called
 the ``Voros coefficient''\cite{Voros:1983,Delabaere:1993} among mathematicians)
is to use consistency relations\cite{Berry:1972na}, which is very
 familiar among physicists.
Although the consistency relation cannot give the phase of the
normalization factor, it is usually sufficient to discuss cosmological
particle production\cite{Kofman:1997yn}.
The exact calculation of the Voros factor
is possible in terms of the EWKB, which is calculated in
Refs.\cite{Voros:1983,Delabaere:1993,Silverstone:2008, Takei:2008,
 Aoki:2009} for the MTP.
(The typical example is shown in
 Fig.\ref{fig-invpot} for the scattering with an inverted quadratic
 potential.)
 and a loop structure of a Bessel-like equation\cite{Aoki:2019}.
While we do not always use these exact calculations explicitly, it is
important to note that the mathematical proofs always support the
calculations in the background.

\begin{figure}[tbp]
\centering
\includegraphics[width=.45\textwidth]{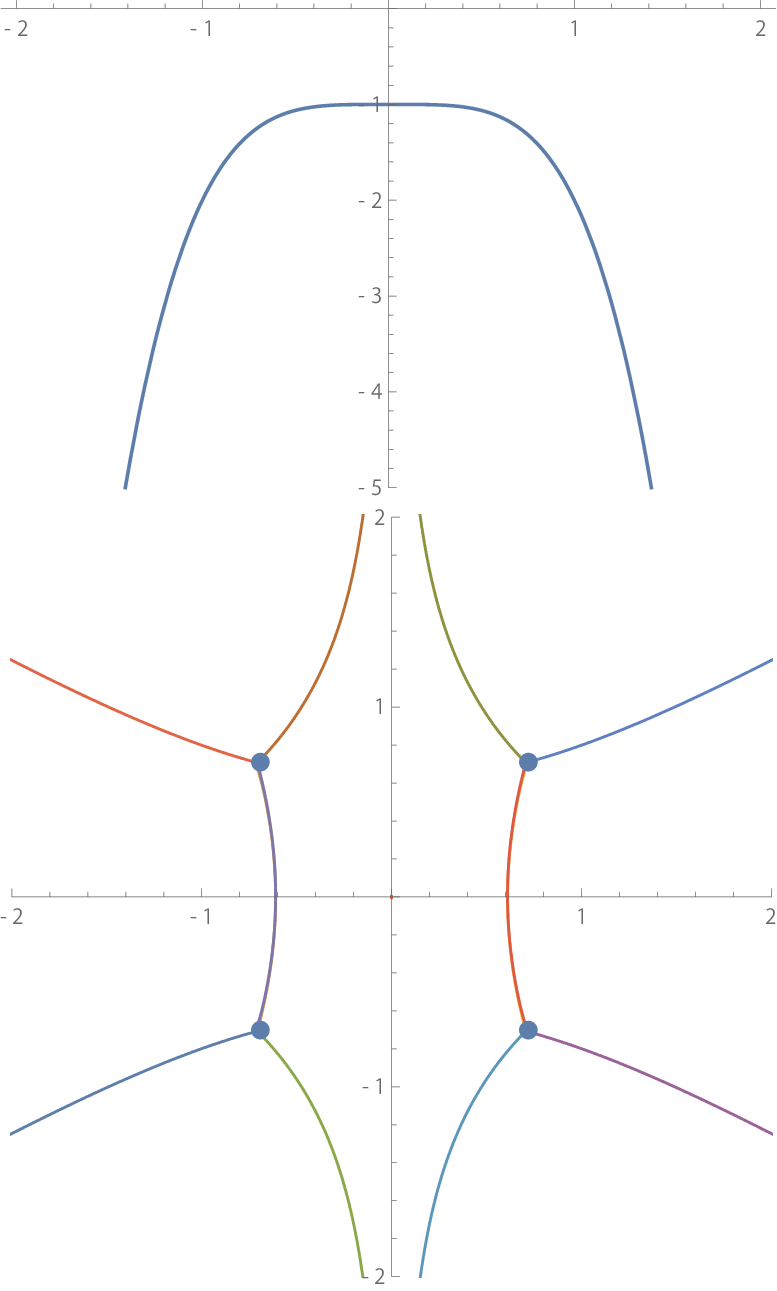}
 \caption{The potential ($V(x)=-1-x^4$ on the $V-x$-plane) is shown in
 the upper panel, and the Stokes lines (for $Q(x)=-1-x^4$ on the complex
 $x$-plane) are shown in the lower panel.
One can see two characteristic structures (called ``MTP'').
The Stokes phenomena (particle production) occur when the real axis of
 time crosses the Stokes line.
In this case, the special function that gives the exact
 solution is not known.
However, from the Stokes lines, one can calculate the connection
 matrices using the EWKB.}
\label{fig-quartic}
\end{figure}

Since the Stokes phenomena of each MTP structure can be calculated 
by the EWKB formalism\cite{Enomoto:2020xlf,Enomoto:2021hfv}, one can easily obtain the local
Bogoliubov transformation occurring at each MTP structure.
Using the EWKB, the parameter $\kappa$ (appeared in
Sec.\ref{subsec-invst} for the Weber function)
is given by\cite{Enomoto:2020xlf}
\begin{eqnarray}
\pi \kappa&=&\int^{x_u}_{x_d} S_{odd} dx,
\end{eqnarray}
where $x_u$ and $x_d$ $(Im[x_u] > Im[x_d])$ are the turning points of
the MTP.
One might be skeptical about the calculation of the above integral
since $S_{odd}$ contains an infinite number of terms.
On the complex $t$-plane, the integration can be replaced by a contour
integral, which can be calculated very easily for the inverted quadratic
potential.
Since the EWKB calculation depends only on the Stokes lines and
$\kappa$, one can easily calculate the connection matrices of 
a more complex $Q(x)$ at the MTP structure without referring to the
special functions.
Indeed, the scattering by a ``quartic'' potential can be
explained by a pair of such MTP structures.
(In this case, $\kappa$ of each MTP has to be calculated by the integral
of the EWKB.)
The situation is shown in Fig.\ref{fig-quartic}.
We hope the figure helps one understand the importance of the
MTP structure.
For mathematical details we refer to Ref.\cite{Voros:1983}.

Here is an obvious example where particle production cannot be explained
by the MTP structure. Scattering by ``tanh''-type potentials gives
infinitely degenerate Stokes lines, whose solutions are given by
hypergeometric functions\cite{Birrell:1982ix, Taya:2020dco}.

In the case of an inverted quadratic potential, $\kappa$ of 
the Weber function is identical to the EWKB integral above. 
In other cases (e.g., the quartic potential in
Fig.\ref{fig-quartic}),  there is no special function that
describes the exact solution, so the Stokes phenomenon for the MTP
structure must be calculated using integrals.

Since interaction usually raises the order of the differential equation,
the problem becomes more complicated than in the original discussion 
of the Schr\"odinger-type equation.
It should be noted that a complete review of the mathematics behind this
analysis is not possible. 
The analysis of the Stokes phenomena for 
higher-order differential equations can be found in
Refs.\cite{Virtual:2015HKT,Aoki:2019,Takei:2008}.
To avoid confusion, we will review the EWKB of higher-order
differential equations after describing the model setup.

\section{Bosonic preheating and the Stokes
 phenomena}
At least one complex scalar field is needed for our 
discussion of matter-antimatter asymmetry.
A closer examination reveals problems with the conventional definition
of the creation and annihilation operators in asymptotic states.
First, we explain why quantization of a complex scalar field requires
special care when introducing CP symmetry breaking.

\subsection{Quantization of a real scalar field}
The Lagrangian density of a real scalar field is
\begin{eqnarray}
{\cal L}=-\frac{1}{2}(\partial_\mu \varphi)(\partial^\mu
 \varphi)-\frac{1}{2}m^2 \varphi^2.
\end{eqnarray}
The quantization is
\begin{eqnarray}
\varphi(x)&=&\int\frac{d^3\mathbf{k}}{(2\pi)^3}\frac{1}{\sqrt{2\omega_\mathbf{k}}}
\left(a_\mathbf{k}e^{-i\omega_\mathbf{k}t}+a^\dagger_\mathbf{-k}e^{i\omega_\mathbf{k}t}
\right)e^{i\mathbf{kx}}\nonumber\\
&=&
\int\frac{d^3\mathbf{k}}{(2\pi)^3}\frac{1}{\sqrt{2\omega_\mathbf{k}}}
\left(a_\mathbf{k}e^{i(-\omega_\mathbf{k}t+\mathbf{kx})}
+a^\dagger_\mathbf{k}e^{i(\omega_\mathbf{k}t-\mathbf{kx})}
\right)\nonumber\\
\pi(x)&=&-i\int\frac{d^3\mathbf{k}}{(2\pi)^3}\sqrt{\frac{\omega_\mathbf{k}}{2}}
\left(a_\mathbf{k}e^{i(-\omega_\mathbf{k}t+\mathbf{kx})}
-a^\dagger_\mathbf{k}e^{i(\omega_\mathbf{k}t-\mathbf{kx})}
\right),\nonumber\\
\end{eqnarray}
where $\omega_\mathbf{k}\equiv\sqrt{\mathbf{k}^2+m^2}$.
Here $\mathbf{x}$ is three-dimensional and $x$ is four-dimensional.

\subsection{Quantization of a complex scalar field}
A free complex scalar field $\phi$ can be written using two independent
real scalar fields $\varphi_1$ and $\varphi_2$;
\begin{eqnarray}
\phi&=&\frac{1}{\sqrt{2}}(\varphi_1+i\varphi_2)\nonumber\\
\pi_\phi&=&\frac{\partial{\cal L}}{\partial\dot{\phi}}=\dot{\bar{\phi}}\nonumber\\
&=&\frac{1}{\sqrt{2}}(\pi_{\varphi_1}-i\pi_{\varphi_2})
\end{eqnarray}
The conventional quantization of the complex scalar field assumes
$\omega_\mathbf{k}\equiv\omega_\mathbf{1k}=\omega_\mathbf{2k}$
to define the creation and the annihilation operators as
\begin{eqnarray}
\label{eq-q-comp}
\phi(x)
&=&
\int\frac{d^3\mathbf{k}}{(2\pi)^3}\frac{1}{\sqrt{2\omega_\mathbf{k}}}
\left(\frac{a_{1\mathbf{k}}+ia_{2\mathbf{k}}}{\sqrt{2}}
e^{i(-\omega_\mathbf{k}t+\mathbf{kx})}
\right.\nonumber\\
&&\left.
+\frac{a^\dagger_{1\mathbf{k}}+ia^\dagger_{2\mathbf{k}}}{\sqrt{2}}
e^{i(\omega_\mathbf{k}t-\mathbf{kx})}
\right)\nonumber\\
&\equiv&
\int\frac{d^3\mathbf{k}}{(2\pi)^3}
\frac{a_{\mathbf{k}}e^{-ikx}+b^\dagger_{\mathbf{k}}e^{ikx}}{\sqrt{2\omega_\mathbf{k}}}\nonumber\\
\bar{\phi}(x)
&=&
\int\frac{d^3\mathbf{k}}{(2\pi)^3}
\frac{b_{\mathbf{k}}e^{-ikx}+a^\dagger_{\mathbf{k}}e^{ikx}}{\sqrt{2\omega_\mathbf{k}}}.
\end{eqnarray}
where in the last result we have used four-dimensional $k$.
In this formalism, the creation and annihilation operators of the
complex scalar field are defined by 
\begin{eqnarray}
\label{eq-trans}
a_\mathbf{k}&=&\frac{a_{1\mathbf{k}}+ia_{2\mathbf{k}}}{\sqrt{2}}\nonumber\\
b_\mathbf{k}&=&\frac{a_{1\mathbf{k}}-ia_{2\mathbf{k}}}{\sqrt{2}}.
\end{eqnarray}
The above definitions are used to define asymptotic states.
We will see why these definitions are causing trouble in our case.

Let us first examine the validity of the primary assumption
$\omega_\mathbf{1k}=\omega_\mathbf{2k}$.
To define the matter and the antimatter states for the asymptotic
states, we consider the case in which CP-violating interactions vanish
at $t=\pm \infty$.
The above definitions of the creation and the
annihilation operators are valid in those limits.
On the other hand, due to the CP-violating interaction, the above states
could be mixed during preheating, and there is no reason to believe
in $\omega_\mathbf{1k}=\omega_\mathbf{2k}$.

To see what happens, we introduce the simplest term of CP violation
as\footnote{Instead of using a complex scalar field, one can introduce
two complex fields $\phi_a$ and $\phi_b$ 
and define the CP violating interaction by 
$\sim [g \Lambda(t)\phi_a\phi_b +
h.c.]$. In this case, the corresponding asymmetry is 
$\Delta n \equiv (n_a+n_b)-(n_{\bar{a}}+n_{\bar{b}})=(n_a-n_{\bar{a}})
+(n_{b}-n_{\bar{b}})\ne 0$.}
\begin{eqnarray}
\label{eq-cpterm}
{\cal L}_{CP}&=&\frac{1}{2}\left[\Lambda(t) \phi^2+h.c\right]\nonumber\\
&=&\frac{1}{2}\left[\Lambda(\varphi_1^2-\varphi_2^2+2i\varphi_1\varphi_2)
+\Lambda^*(\varphi_1^2-\varphi_2^2-2i\varphi_1\varphi_2)\right]\nonumber\\
&=&\Lambda_R(\varphi_1^2 -\varphi_2^2)-2\Lambda_I\varphi_1\varphi_2,
\end{eqnarray}
where we defined $\Lambda=\Lambda_R+i\Lambda_I$.
Now we consider a situation where $\Lambda_R$ is a constant and the
function $\Lambda_I(t)$ can be regarded as a perturbation parameter.
Expanding $\varphi_i=\varphi_{i(0)}+\varphi_{i(1)}+...$,
the perturbative calculation with the time-dependent interaction 
$\Lambda_I(t)$ gives
\begin{eqnarray}
(\partial_t+\mathbf{k}^2+m_\phi^2-2\Lambda_R)\varphi_{1(1)}&=&\Lambda_I(t)\varphi_{2(0)}\nonumber\\
&=&\Lambda_I(t)e^{-i\omega_\mathbf{2}t},
\end{eqnarray}
where (because of ${\cal L}_{CP}$)
\begin{eqnarray}
\omega_1&=&\sqrt{\mathbf{k}^2+m_\phi^2-2\Lambda_R}\nonumber\\
\omega_2&=&\sqrt{\mathbf{k}^2+m_\phi^2+2\Lambda_R}.
\end{eqnarray}
Here the subscript $\mathbf{k}$ is omitted for simplicity.
Using the Green function and defining the Fourier transformation of
$\Lambda_I(t)$ by $\tilde{\Lambda}_I(\omega)$, we have
\begin{eqnarray}
\varphi_{1(1)}=-\int\frac{d\omega'}{2\pi}
\frac{\tilde{\Lambda}_I
(\omega'-\omega_2)}{(\omega')^2-(\mathbf{k}+m_\phi^2-2\Lambda_R)}e^{-i\omega't}.
\end{eqnarray}
Considering the residues at 
$\omega'=- \omega_1$, we find the coefficient of $e^{+ i\omega_1 t}$ to be
\begin{eqnarray}
\beta_{1,\mathbf{k}}=i\frac{\tilde{\Lambda}_I^*(\omega_1+\omega_2)}{2\omega_1}.
\end{eqnarray}
The above calculation is almost the same as the calculation of
Ref.\cite{Dolgov:1996qq} except for the quantization, which is crucial
in our case.
Because of the different signs in front of $\varphi_1^2$ and
$\varphi_2^2$ in 
Eq.(\ref{eq-cpterm}), we have $\omega_1\ne \omega_2$.
This discriminates $\beta_\mathbf{1k}$ and $\beta_\mathbf{2k}$, and could be a 
crucial source of the asymmetry (the mechanism of the asymmetry
generation is not trivial, as we will see below).
What we consider in this paper is the CP-violating interaction, which
introduces intermediate states of $\omega_1\ne\omega_2$.

One can write the mass matrix and the CP-violating interactions in a
matrix form
\begin{eqnarray}
M&=&\left(
\begin{array}{cc}
m_\phi^2& \Lambda^*\\
\Lambda&m_\phi^2
\end{array}
\right),
\end{eqnarray}
which normally gives the higher-order differential equations
after decoupling.
Note however the matrix can be diagonalized and the equations become the
second-order differential equations if all elements are
constant.
Using the formalism of the exact WKB for the higher-order
differential equations, we are going to find the source of asymmetry
in this model.

\subsection{Case1: Real $\Lambda(t)$}
Let us first assume that $\Lambda(t)$ is a real function.
Using Eq.(\ref{eq-cpterm}) and $\Lambda_I=0$, the interaction between
$\varphi_1$ and $\varphi_2$ vanishes, and the equations of
$\varphi_1,\varphi_2$ are separated.
In this case, we do not have to consider the higher-order differential equations.
On the other hand, their masses are distinguished as 
\begin{eqnarray}
m_1^2&=&m_\phi^2-2\Lambda_R\nonumber\\
m_2^2&=&m_\phi^2+2\Lambda_R.
\end{eqnarray}
For the real  $\Lambda(t)$, $\varphi_1$ and $\varphi_2$ are
independent and are not equal.
The Stokes phenomenon occurs independently and does not mix 
the asymptotic solutions of $\varphi_1$ and $\varphi_2$.
Our question is whether the asymmetry production is possible or not in this
case.
To be more precise, we are going to identify the Stokes
phenomena, which is responsible for the matter-antimatter asymmetry.
What we can assume here is just $\beta_\mathbf{11k}\ne
\beta_\mathbf{22k}$ and $\beta_\mathbf{12k}=\beta_\mathbf{21k}=0$.
(The subscripts $\mathbf{12k}$ and $\mathbf{21k}$ denote the mixing of
solutions between $\varphi_1$ and $\varphi_2$.)
Does this difference source the matter-antimatter asymmetry in the
asymptotic states?
To understand the situation, we will write the Bogoliubov transformation of
the original creation and annihilation operators in
a matrix form and translate it to the asymptotic states.
For the relation between the asymptotic states of Eq.(\ref{eq-trans})
and the original creation and annihilation operators of $\varphi_1$ and
$\varphi_2$, we define in the limit $t=-\infty$ as
\begin{eqnarray}
\label{eq-trans2}
\left(
 \begin{array}{c}
  a_\mathbf{k}\\
  a^\dagger_\mathbf{k}\\
  b_\mathbf{k}\\
  b^\dagger_\mathbf{k}
 \end{array}
\right)&=&\frac{1}{\sqrt{2}}
\left(
 \begin{array}{cccc}
  1&0&i&0\\
  0&1&0&-i\\ 
  1&0&-i&0\\
  0&1&0&i
 \end{array}
\right)
\left(
 \begin{array}{c}
  a_{1\mathbf{k}}\\
  a^\dagger_{1\mathbf{k}}\\
  a_{2\mathbf{k}}\\
  a^\dagger_{2\mathbf{k}}
 \end{array}
\right).
\end{eqnarray}
We can write the Bogoliubov transformations of the original operators
in the following form;
\begin{eqnarray}
\label{eq-bogo1}
\left(
 \begin{array}{c}
  a_{1\mathbf{k}}\\
  a^\dagger_{1\mathbf{k}}\\
  a_{2\mathbf{k}}\\
  a^\dagger_{2\mathbf{k}}
 \end{array}
\right)'
&=&
\left(
 \begin{array}{cccc}
  \alpha_\mathbf{11k}&\beta_\mathbf{11k}&0&0\\
  \beta^*_\mathbf{11k}&\alpha^*_\mathbf{11k}&0&0\\ 
  0&0&\alpha_\mathbf{22k}&\beta_\mathbf{22k}\\
  0&0&\beta^*_\mathbf{22k}&\alpha^*_\mathbf{22k}\\ 
 \end{array}
\right)
\left(
 \begin{array}{c}
  a_{1\mathbf{k}}\\
  a^\dagger_{1\mathbf{k}}\\
  a_{2\mathbf{k}}\\
  a^\dagger_{2\mathbf{k}}
 \end{array}
\right)\nonumber\\
\end{eqnarray}
Using Eq.(\ref{eq-trans2}) and Eq.(\ref{eq-bogo1}), the Bogoliubov
transformations of the asymptotic states are given by 
\begin{eqnarray}
a_\mathbf{k}&\rightarrow&\frac{1}{\sqrt{2}}\left(
 \alpha_\mathbf{11k}a_\mathbf{1k}
 +i\alpha_\mathbf{22k}a_\mathbf{1k}
 +\beta_\mathbf{11k}a^\dagger_\mathbf{1k} 
 +i\beta_\mathbf{22k}a^\dagger_\mathbf{2k}\right)\nonumber\\
b_\mathbf{k}&\rightarrow&\frac{1}{\sqrt{2}}\left(
 \alpha_\mathbf{11k}a_\mathbf{1k}
 -i\alpha_\mathbf{22k}a_\mathbf{1k}
 +\beta_\mathbf{11k}a^\dagger_\mathbf{1k}
-i\beta_\mathbf{22k}a^\dagger_\mathbf{2k}\right),
\end{eqnarray}
which give
\begin{eqnarray}
n=\bar{n}&=&\frac{|\beta_\mathbf{11k}|^2+|\beta_\mathbf{22k}|^2}{2}.
\end{eqnarray}
Therefore, despite the difference between $\varphi_1$ and
$\varphi_2$, asymmetry production is impossible for real $\Lambda(t)$.
A possible source of the asymmetry is the contribution from the
off-diagonal blocks ($\alpha_\mathbf{12k}$ and $\beta_\mathbf{12k}$,
which are $0$ in Eq.(\ref{eq-bogo1})), which could introduce the asymmetry as
$|\beta_\mathbf{11k}+i\beta_\mathbf{12k}|^2\ne|\beta_\mathbf{11k}-i\beta_\mathbf{12k}|^2$.
This is nothing but a quantum interference between different Stokes phenomena.
However, the off-diagonal blocks always vanish for the real $\Lambda(t)$.
We are going to examine this possibility in the next section.

\subsection{Case2: Complex $\Lambda(t)$}
We have learned that the Stokes phenomena, which do not mix $\varphi_1$
and $\varphi_2$, cannot generate the required asymmetry even though 
$\varphi_1$ and $\varphi_2$ are distinguishable.
The asymmetry can be generated if the interference
$|\beta_\mathbf{11k}+i\beta_\mathbf{12k}|^2\ne|\beta_\mathbf{11k}-i\beta_\mathbf{12k}|^2$
is possible, but what process is responsible for the
off-diagonal blocks in the matrix of Eq.(\ref{eq-bogo1})?

To generate the interference, the Stokes phenomenon has to happen at least
twice.
They are $\beta_\mathbf{11k}$ or $\beta_\mathbf{22k}$ ($\varphi_1$ and $\varphi_2$ are
not mixed) and $\beta_\mathbf{12k}$(mixed). 
Therefore, we need a higher-order differential equation to realize the Stokes
phenomena.
Since the higher-order differential equation is very complicated,
we are going to pick up the essentials and explain how to introduce the
asymmetry in physics.

\section{The EWKB for the higher-order differential equations}
In this section, we discuss the Stokes phenomena for a complex scalar
field with a CP-violating interaction.
We begin with the simple cases of second and third-order differential
equations to illuminate the Stokes phenomena of the higher-order
differential equations. 
Finally, we show the local structure of the Stokes lines that cause
the asymmetry. 

\subsection{From the first-order simultaneous differential
  equation to the Schr\"odinger equation}
Before discussing the simultaneous second-order differential
equation with a $2\times 2$ matrix, 
let us first discuss a simultaneous first-order differential equation with
a $2\times 2$ matrix.
The point of discussion is the relationship between the conventional
Schr\"odinger-like equation (where the solutions are given by simple $\pm$
solutions) and the formulation given by $\zeta$.
(What $\zeta$ is will be explained in this section.) 

In cosmology, a simultaneous first-order differential equation with
a $2\times 2$ matrix is found in the dynamical particle production 
of Majorana fermions, which is very similar to the Landau-Zener
model.
(The Landau-Zener model is normally discussed for constant off-diagonal
elements. See Appendix \ref{app-zener} for more details.)
Following Refs.\cite{Enomoto:2020xlf,Enomoto:2021hfv}, we start 
with\footnote{Fermionic preheating of 
the Majorana fermion has been discussed in Ref.\cite{Enomoto:2020xlf,
Enomoto:2021hfv} in detail comparing the conventional special functions(the Weber
function) and the EWKB.}
\begin{eqnarray}
\label{eq-simpleoriginalLZ}
i\hbar \frac{d}{dt}\left(
\begin{array}{c}
X\\
Y
\end{array}
\right)&=&\left(
\begin{array}{cc}
D(t) & \Delta(t)^*\\
\Delta(t) & -D(t)
\end{array}
\right)
\left(
\begin{array}{c}
X\\
Y
\end{array}
\right),
\end{eqnarray}
which can be decoupled to give 
\begin{eqnarray}
\hbar^2 \ddot{X}-\hbar\frac{\dot{\Delta}^*}{\Delta^*}\dot{X}+
\left(-\frac{i\hbar D\dot{\Delta}^*}{\Delta^*}
+i\hbar \dot{D}
+|\Delta|^2+D^2
\right)X=0.\nonumber\\
\end{eqnarray}
We introduce $p_1(\hbar,t)$ and $p_0(\hbar,t)$ to rewrite the equation in the following
form:
\begin{eqnarray}
\label{eq-orig2nd}
\ddot{X}+\frac{p_1(\hbar,t)}{\hbar}\dot{X}+\frac{p_0(\hbar,t)}{\hbar^2}X&=&0.
\end{eqnarray}
The ``Schr\"odinger equation'' can be recovered by using the transformation
\begin{eqnarray}
\label{eq-transb}
X&=&\exp\left(-\frac{1}{2}\int^x \frac{p_1}{\hbar} dx\right)\hat{X}
\end{eqnarray}
Finally, we have the second-order ordinary equation of the
Schr\"odinger-type;
\begin{eqnarray}
\label{Eq_2ndorder-Majorana}
\hbar^2\ddot{\hat{X}}&+&\left(p_0-\frac{p_1^2}{4}-\frac{\hbar}{2}p_1'\right)\hat{X}=0.
\end{eqnarray}
Applying the EWKB, $\hat{S}_{-1}$ becomes 
\begin{eqnarray}
\hat{S}_{-1}^2&=&-\left(|\Delta|^2+D^2\right)+\frac{(\dot{\Delta})^2}{4(\Delta)^2},
\end{eqnarray}
which gives the $\pm$-solutions.

In the above calculation, we have introduced a non-trivial
transformation to find the Schr\"odinger-type equation.
Alternatively, for the original $X$ of Eq.(\ref{eq-orig2nd}), one may calculate
\begin{eqnarray}
S_{-1}^2+p^{(0)}_1(t)S_{-1}+p^{(0)}_0(t)&=&0,
\end{eqnarray}
where $p_i^{(n)}$ is the coefficient of $\hbar^n$ when $p_i$ is expanded
by $\hbar$.
For $S_{-1}$ of the original equation, we have two solutions
\begin{eqnarray}
\zeta_\pm&=&-\frac{p_1^{(0)}\pm \sqrt{(p_1^{(0)})^2-4 p_0^{(0)}}}{2},
\end{eqnarray}
which define the ``turning points'' at the solutions of
$\zeta_+=\zeta_-$.
In this case, the Stokes lines are defined by
\begin{eqnarray}
\label{eq-stokeshigher}
\mathrm{Im}\left[\int_a^t (\zeta_+-\zeta_-)dt \right]&=&0,
\end{eqnarray}
where the constant $a$ normally denotes the turning point from which the
Stokes lines are coming out.

The above formulation of $\zeta_\pm$ is more general than
the formulation based on the Schr\"odinger-type equation.
It is easy to see that the $\zeta$-formalism is essential to the EWKB
for higher-order differential equations.
Also, when applied to the Landau-Zener transition at ``level
crossings'', the $\zeta$-formalism is much more intuitive than 
the Schr\"odinger-type formalism.

From the above discussion, it is easily understood that
Schr\"odinger-type equations can be reformed from $\zeta_\pm$.
Although not accurate, (see Ref.\cite{Virtual:2015HKT} for more accurate
discussions) such reconstruction is also locally possible near a
turning point defined for a pair of solutions 
of higher-order differential equations.
Since in physics of particle creation the Stokes phenomenon 
(which is defined for a pair of solutions) occurs when the
real axis crosses the Stokes line, the Schr\"odinger-type equation
reconstructed from the pair is
responsible for the connection matrix of the particle creation at that moment.
Also, if the linear approximation is possible for the
``level crossings'', the Stokes lines are locally given by the
MTP. 
(See appendix \ref{app-zener} for the ``linear approximation'' used in
the original Landau-Zener model.)
Since the Stokes phenomena of the MTP structure is calculable, 
the connection matrix for such a general case can be calculated.
Therefore, although the mathematics of higher-order differential
equations is theoretically very complex, by decomposing the essential
elements into familiar MTP structures, one can find how the 
physics can be embedded.
This is the basic strategy of this paper.

In the next section, we analyze two typical third-order differential
equations in which the new Stokes line appears.

\subsection{The ``new Stokes lines'' of the third-order equation}
\begin{figure}[t]
\centering
\includegraphics[width=.45\textwidth]{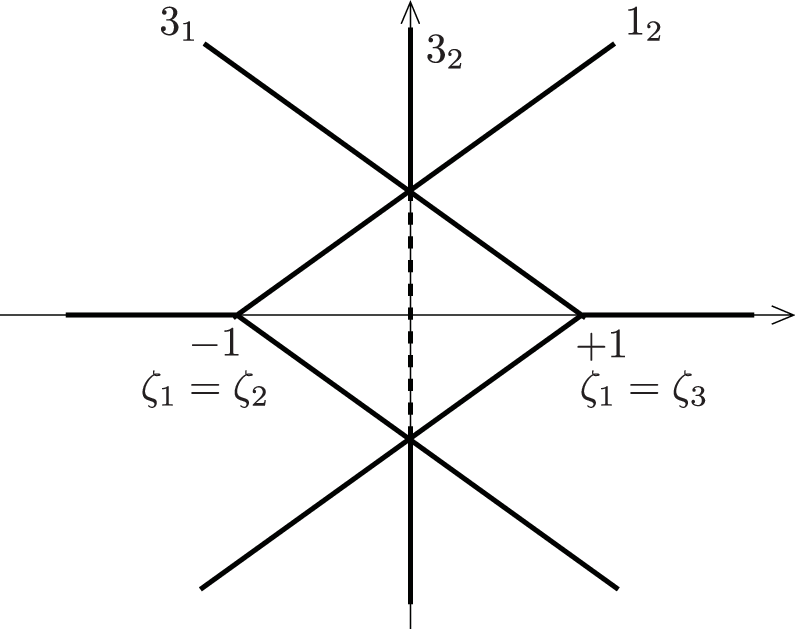}
 \caption{Each Stokes lines are for a pair of two solutions. 
The numberings are denoting the two solutions.
For $3_1$, the dominant solution is ``$3$'', and the Stokes phenomena
 occurs for ``$3$'' and ``$1$''.
The line denoted by $3_2$ is the ``new Stokes line'', but the dotted
 part does not cause the Stokes phenomena. }
\label{fig-NBR}
\end{figure}

Let us extend the EWKB analysis and the formulation of the previous section to
  the third-order ordinary differential equation.
We consider the model given by Berk, Nevis, and Roberts(NBR)\cite{NBR},
  which is defined by
\begin{eqnarray}
\left[
\frac{d^3}{dt^3}+3\eta^2\frac{d}{dt}+2i t \eta^3
\right]\psi(t)&=&0,
\end{eqnarray}
where $\eta\equiv\hbar^{-1}$ is a large parameter for the WKB expansion.
$S_{-1}$ obeys the equation given by
\begin{eqnarray}
\zeta^3+3\zeta+2it&=&0,
\end{eqnarray}
which has solutions 
\begin{eqnarray}
\zeta_1&=&-\frac{1}{\left(-i t + \sqrt{1 - t^2}\right)^\frac{1}{3}} + \left(-i t + \sqrt{1 -
 t^2}\right)^\frac{1}{3}\nonumber\\
\zeta_2&=&
\frac{1+i \sqrt{3}}{2 \left(\sqrt{1-t^2}-i t\right)^\frac{1}{3}}-\frac{1}{2} \left(1-i
							       \sqrt{3}\right)
\left(\sqrt{1-t^2}-i t\right)^\frac{1}{3}\nonumber\\
\zeta_3&=&
\frac{1-i \sqrt{3}}{2 \left(\sqrt{1-t^2}-i t\right)^\frac{1}{3}}-\frac{1}{2} \left(1+i
							       \sqrt{3}\right)
\left(\sqrt{1-t^2}-i t\right)^\frac{1}{3}.\nonumber\\
\end{eqnarray}
The formal turning points are $t=-1$ for $\zeta_1=\zeta_2$ and
$t=1$ for $\zeta_1=\zeta_3$, both of which are the Airy-type (i.e, three
Stokes lines are coming out of a turning point).
The Stokes lines are presented in Fig.\ref{fig-NBR}.

Without the ``new Stokes line'' denoted by $3_2$ in Fig.\ref{fig-NBR}, 
the connection formulae are inconsistent between the route above and
below the crossing points of $1_2$ and $3_1$.
(Details of the calculation are given in Ref.\cite{NBR}. Without the New
Stokes line, the connection matrix seems to depend on the route.)
The new Stokes line ($3_2$) was introduced to solve the problem of
the global inconsistency.
Note that the Stokes phenomenon is not observed on the dotted part.
(This was the speculation in Ref.\cite{NBR}, but later
works\cite{Aoki:2019, Takei:2008, Virtual:2015HKT} revealed the
mathematics behind it.)
The regular and the dotted parts of the new Stokes line are easily 
distinguished by considering the consistency of the Stokes phenomena
around the crossing points.

Using the EWKB, one can find an extra turning point at the origin, which
was named the ``virtual turning point''\cite{Virtual:2015HKT}.
The virtual turning point is the solution of
\begin{eqnarray}
\int^{t_{v}}_{t_{12}}(\zeta_2-\zeta_1)dt&=&\int^{t_{v}}_{t_{13}}(\zeta_3-\zeta_1)dt.
\end{eqnarray}
Here $t_{ij}$ is the solution of $\zeta_i=\zeta_j$ (normal turning
points), which are $t_{12}=-1$ and $t_{13}=1$ in this case.
One can solve the equation to find the virtual turning point at
$t_v=0$.
Since the virtual turning point is placed on the dotted line, it causes
nothing in physics (but is very important in mathematics).

In the above example of the NBR equation, there were two Airy-type 
turning points on the real axis.
Although the NBR model is very informative and historically important
for understanding the new Stokes lines,
it does not describe the scattering problem in physics. 
Therefore, we will consider another model in which the Stokes
lines have local MTP structures.
Following Ref.\cite{Virtual:2015HKT}, we consider a three-level
Landau-Zener model\cite{Zener:1932ws}.
In the multi-level Landau-Zener model, the MTP structure appears at each
level crossing.
In the three-level Landau-Zener model, there are three kinds of
 MTP for level crossings because level crossings are defined for
 pairs.

The model is described by
\begin{eqnarray}
i \frac{d}{dt}\psi&=& \eta H(t,\eta)\psi
\end{eqnarray}
with
\begin{eqnarray}
H(t,\eta)&=&H_0(t)+\eta^{-\frac{1}{2}}H_{\frac{1}{2}}\nonumber\\
H_0(t)&=&\left(
\begin{array}{ccc}
\rho_1(t)& 0&0\\
0& \rho_2(t)& 0\\
0&0&\rho_3(t)
\end{array}
\right)\nonumber\\
&=&\left(
\begin{array}{ccc}
t +3 & 0&0\\
0&  2t& 0\\
0&0& 4 t
\end{array}
\right)\nonumber\\
H_\frac{1}{2}&=&\left(
\begin{array}{ccc}
0& c_{12}&c_{13}\\
\bar{c}_{12}& 0& c_{23}\\
\bar{c}_{13}&\bar{c}_{23}&0
\end{array}
\right).
\end{eqnarray}
After decoupling, $S_{-1}$ obeys the equation 
\begin{eqnarray}
\zeta^3+i(7t+3) \zeta^2 -2t(7t+9)\zeta
-8it^2(t+3)&=&0,
\end{eqnarray}
which has three solutions given by
\begin{eqnarray}
\label{eq-3LZsol}
\zeta_1=&i(t+3)&=i \rho_1(t)\nonumber\\
\zeta_2=&i 2t&=i\rho_2(t)\nonumber\\
\zeta_3=&i 4t&=i\rho_3(t).
\end{eqnarray}
The turning points are
\begin{eqnarray}
\zeta_2=\zeta_3&\rightarrow&t_{23}=0\nonumber\\
\zeta_1=\zeta_3&\rightarrow&t_{13}=1\nonumber\\
\zeta_1=\zeta_2&\rightarrow&t_{12}=3,
\end{eqnarray}
where $t_{23}<t_{13}<t_{12}$.

From this result, one will understand how the off-diagonal elements of
the above model simplify the argument.
Because of the factor $\eta^{-\frac{1}{2}}$ in front of
$H_{\frac{1}{2}}$, elements ($c_{ij}$) disappear from $\zeta_i$.
Although extra $\hbar$ could not be acceptable in physics,
this factor mimics small interaction and was introduced to make the
model instructive.
In this model, the MTP structure of the Landau-Zener model is the double
turning point, from which the four Stokes lines extend.

\begin{figure}[t]
\centering
\includegraphics[width=.45\textwidth]{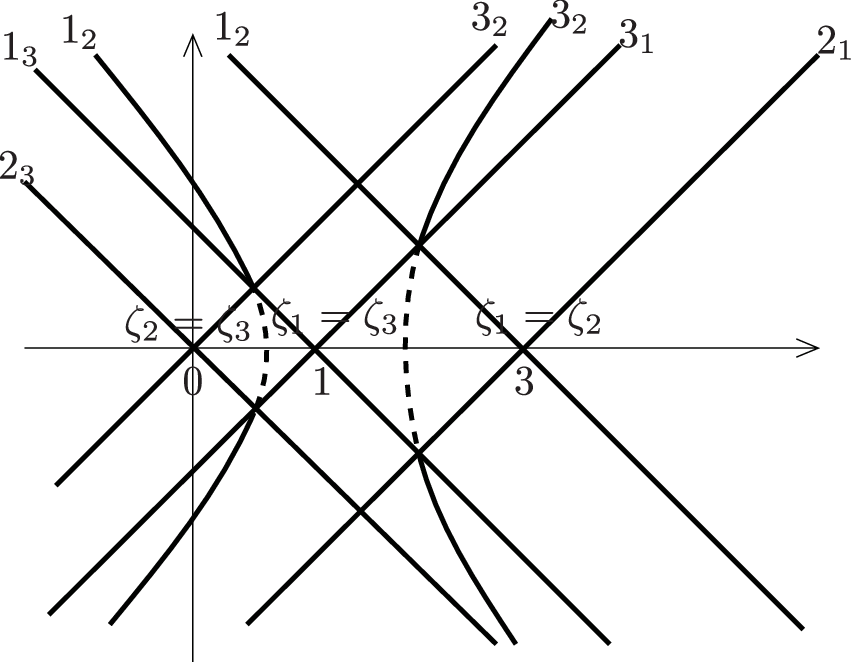}
 \caption{Unlike the NBR model, each level crossing gives 
the (degenerated) MTP structure, from which four Stokes lines are coming
 out.}
\label{fig-NBR2}
\end{figure}

In this model, turning points are found at $t=0,1,3$ and the structure of
the Stokes lines is very simple, as is shown in Fig.\ref{fig-NBR2}.
One can see ``ordered crossing points''(a new Stokes line is needed)
and ``non-ordered crossing points''(a new Stokes line is not needed) in
Fig.\ref{fig-NBR2}.
To understand the ``consistency'' required for the Stokes lines, 
one can explicitly write down the connection matrices of the Stokes
phenomena.
At the crossing point of $1_2$ and $3_2$, the Stokes phenomena do not
depend on the route because the matrices $U_{1_2}$ and $U_{3_2}$
commute, while at the crossing point of $1_3$ and $3_2$, the matrices 
$U_{1_3}$ and $U_{3_2}$ do not commute.
To make the Stokes phenomena route-independent, one has to introduce
another stokes line at the crossing point of $1_3$ and $3_2$.
Of course, the mathematics behind this phenomenon is not so trivial.
More details and rigorous arguments are described in
Ref.\cite{Virtual:2015HKT}.

\subsection{The EWKB for a complex scalar field with CP violation}
Finally, we are going to solve the equations for the real scalar fields
$\varphi_{1,2}$ of the complex scalar $\phi$.
We write
\begin{eqnarray}
-\hbar^2 \frac{d^2}{dt^2}\left(
\begin{array}{c}
\varphi_1\\
\varphi_2
\end{array}
\right)&=&
\Omega(t)\left(
\begin{array}{c}
\varphi_1\\
\varphi_2
\end{array}
\right),
\end{eqnarray}
where the matrix $\Omega(t)$ is given by
\begin{eqnarray}
\Omega(t)&=&
\left(
\begin{array}{cc}
D_1& \Delta\\
\Delta& D_2
\end{array}
\right)\nonumber\\
&=&
\left(
\begin{array}{cc}
\omega_\mathbf{k1}^2-2\Lambda_R& 2\Lambda_I\\
2 \Lambda_I& \omega_\mathbf{k2}^2+2\Lambda_R
\end{array}
\right).
\end{eqnarray}
After separating $\Lambda(t)=\Lambda_R+i\Lambda_I$, all elements of the
matrix are real.
Defining
\begin{eqnarray}
\varphi_1^{(i)}\equiv \frac{d^i}{dt^i}\varphi_1,
\end{eqnarray}
the decoupled equations (the 4th-order ordinary differential equations) can be written as
\begin{eqnarray}
\sum_{i=0}^4 p_i \varphi_1^{(i)}&=&p_4 \varphi_1^{(4)}+p_3
 \varphi_1^{(3)}+p_2 \varphi_1^{(2)}+p_1 \varphi_1^{(1)}+p_0
 \varphi_1\nonumber\\ 
&=&0.
\end{eqnarray}
Coefficients are given by
\begin{eqnarray}
\frac{p_4}{\hbar^4}&=&1\nonumber\\
\frac{p_3}{\hbar^3}&=&-2\hbar\frac{\Delta'}{\Delta}\nonumber\\
\frac{p_2}{\hbar^2}&=&D_1+D_2
+\hbar^2\left[\frac{2(\Delta')^2-\Delta\Delta''}{\Delta^2}\right]\nonumber\\
\frac{p_1}{\hbar}&=&2\hbar\frac{-D_1\Delta'+\Delta D_1'}{\Delta}\nonumber\\
p_0&=&-\Delta^2+D_1D_2+\hbar^2\left[-\frac{2 \Delta' D_1'-2D_1(\Delta')^2}{\Delta}
+D_1''-D_1\Delta''\right].\nonumber\\
\end{eqnarray}
Assuming that time-dependent background fields are all external,
we have 
\begin{eqnarray}
\zeta^4 +2  (D_1+D_2) \zeta^2 +
D_1D_2 -\Delta^2&=&0,
\end{eqnarray}
which has four solutions given by
\begin{eqnarray}
\zeta_{A\pm}&\equiv&\pm\frac{\sqrt{-(D_1+D_2)+\sqrt{4
 \Delta^2+(D_1-D_2)^2}}}{\sqrt{2}}\nonumber\\
&=&\pm\frac{\sqrt{-(\omega_\mathbf{k1}^2+\omega_\mathbf{k2}^2)
+4\sqrt{\Lambda_I^2+\Lambda_R^2}}}{\sqrt{2}}\nonumber\\
\zeta_{B\pm}&\equiv&\pm\frac{\sqrt{-(D_1+D_2)-\sqrt{4
 \Delta^2+(D_1-D_2)^2}}}{\sqrt{2}}\nonumber\\
&=&\pm\frac{\sqrt{-(\omega_\mathbf{k1}^2+\omega_\mathbf{k2}^2)
-4\sqrt{\Lambda_I^2+\Lambda_R^2}}}{\sqrt{2}}.
\end{eqnarray}
Here $\pm$-pair solutions correspond to 
asymptotic $\pm$-solutions of real scalar fields ($\varphi_{1,2}$).

Let us assume that $m_\phi(t), \Lambda_R(t),\Lambda_I(t)$ are all
independent and time-dependent.
(Although the original $m_\phi(t), \Lambda_R(t),\Lambda_I(t)$
are defined to be real, they are complex on the complex $t$-plane.)
The turning points are given by
\begin{enumerate}
\item $\zeta_{A+}-\zeta_{A-}=0$ ($\zeta_{A\pm}=0$);
\begin{eqnarray}
D_1+D_2&=&\sqrt{\Lambda_R^2+\Lambda_I^2}
\end{eqnarray}
\item $\zeta_{B+}-\zeta_{B-}=0$ ($\zeta_{B\pm}=0$);
\begin{eqnarray}
D_1+D_2&=&-\sqrt{\Lambda_R^2+\Lambda_I^2}
\end{eqnarray}
\item $\zeta_{A\pm}-\zeta_{B\pm}=0$;
\begin{eqnarray}
\Lambda_R^2+\Lambda_I^2&=&0,
\end{eqnarray}
\item $\zeta_{A\pm}-\zeta_{B\mp}=0$;
\begin{eqnarray}
\Lambda_R^2+\Lambda_I^2=0&,& D_1+D_2=0.
\end{eqnarray}
\end{enumerate}
The MTP of the last turning points (4.) is expected to generate
$\beta_\mathbf{12k}$ by itself.
However, they can be ignored in physics because the two conditions
must be satisfied simultaneously.
The required $\beta_\mathbf{12k}\ne 0$ will be generated by the
combinations of the Stokes phenomena of the other turning points.
In this section, we will denote the above turning points (1., 2. and 3.)
by the numbers given above. 

In addition, one can see that if the interaction is
given by $\Lambda(t)=\Lambda_0 e^{i\theta(t)}$, it gives
\begin{eqnarray}
\Lambda_I^2+\Lambda_R^2=\Lambda_0^2,
\end{eqnarray}
which extinguishes the required turning points (3.) from the scenario.
Therefore, the asymmetry production is theoretically impossible for 
$\Lambda(t)=\Lambda_0 e^{i\theta(t)}$.
On the other hand, one might be tempted to expand it for $\theta(t)=A \cos
\omega_I t, A \ll 1$ as
\begin{eqnarray}
\Lambda(t)&=&\Lambda_0 e^{i\theta(t)}\nonumber\\
&\simeq& \Lambda_0\left[1+i\theta(t)\right]\nonumber\\
&=&\Lambda_0\left[1+iA\cos \omega_I t\right],
\end{eqnarray}
which can generate false turning points (3.) on the complex $t$-plane.
Note that the above expansion is nothing but the expansion 
considered in Ref.\cite{Dolgov:1996qq} for baryogenesis after natural
inflation.
The above simple calculation shows why naive expansions
are dangerous in non-perturbative analysis.

\begin{figure}[t]
\centering
\includegraphics[width=.45\textwidth]{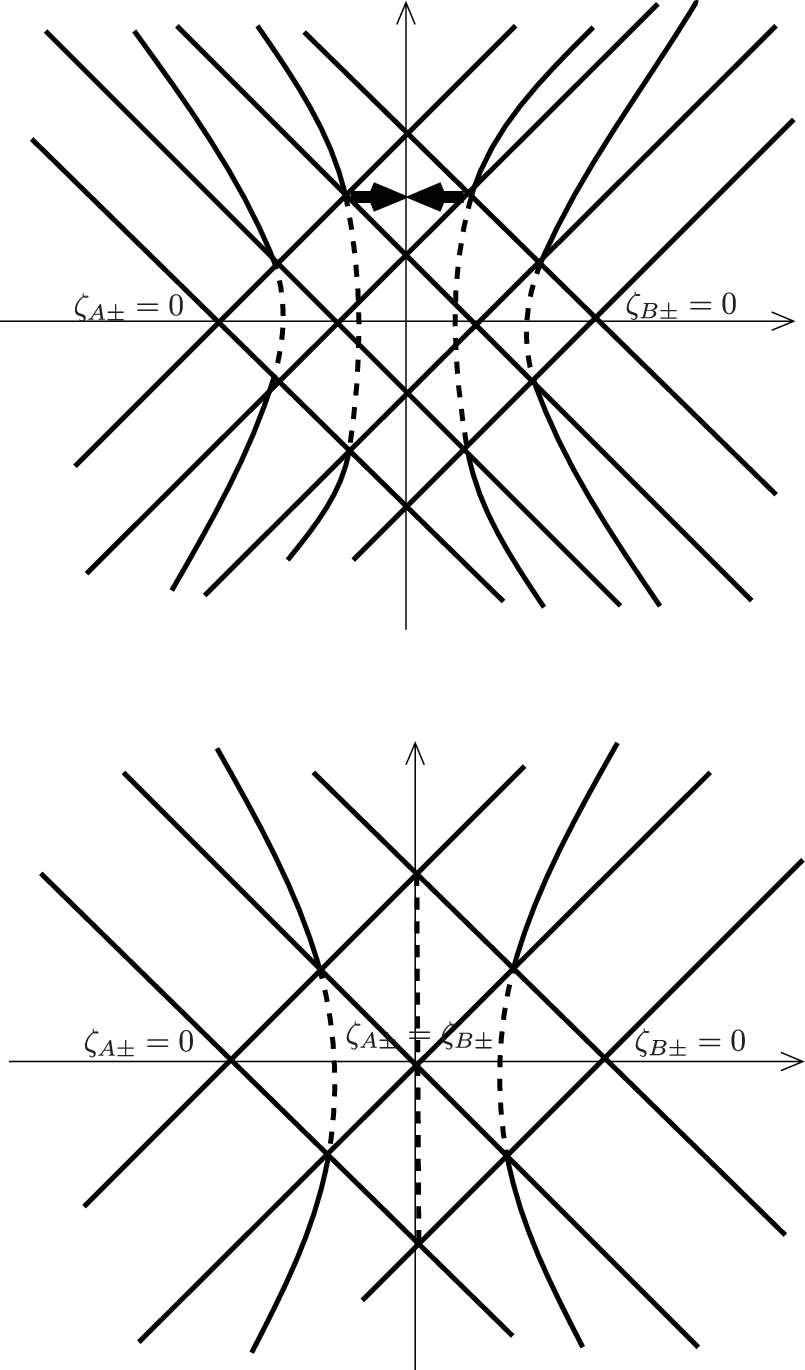}
 \caption{The turning point of 3. is placed between the two (1. and 2.)
 turning points. In the upper panel, the degenerated turning points of
 3. are manually separated for illustration.
For simplicity, each turning point is shown by a double turning
 point of a pair of MTP. 
In the lower panel, mixing between $A$ and $B$ occurs
 near the origin.}
\label{fig-scalarVTP}
\end{figure}

To understand the required structure of the Stokes lines, we show 
the simplest Stokes lines in Fig.\ref{fig-scalarVTP}, for which 
fig.\ref{fig-NBR2} is extended to the four-level Landau-Zener transition.
(Here we have four solutions $\zeta_{A\pm}$ and $\zeta_{B\pm}$.)
For illustration, turning points are manually aligned, and 
are shown to have degenerated MTP structure (i.e, the off-diagonal
elements of the local Landau-Zener transition 
are supposed to be small).
What we are considering here is a simple situation in which
the interference between the Stokes phenomena is possible.
If the particle production is caused by oscillation, there should be
many turning points (MTP) in the global structure.
One can easily construct many exceptional cases (e.g, using tanh to make
the Stokes lines infinitely degenerated, or introducing singularities).

Since the Stokes lines emanating from turning points 1. and 2
do not have a common solution, their intersection obviously does not require
a new Stokes line.
On the other hand, the Stokes lines from the turning point 3. 
have a common solution with the Stokes lines from 1. and 2., which means that
new Stokes lines may emerge from the intersection.

Fig.\ref{fig-scalarVTP} shows that the new Stokes lines are all drawn as
 dotted lines when they intersect the real axis, with no extra influence. 
If the motion of the time-dependent parameters is oscillatory, the MTP
structures of the turning points 1.2. and 3. may appear repeatedly.
However, seeing the Stokes lines of the simplest case, one would expect
that changing the ordering of the turning points will not drastically
change the situation.
To avoid confusion, we confess that ``unfortunately'' we could not find
 out a way to include such effects in physics.
The ``extra'' contribution
from the new Stokes lines, if it had contributed, could have introduced 
the desired interference in a new way. 
We failed to include the new Stokes lines in physics, but our
conclusion is far from a no-go theorem.
Higher-order differential equations are very complex, and what we
have shown in this paper should be only a small part of the whole.
Indeed, for the model considered in Ref.\cite{Shudo:2020}, it was found
 that the new Stokes line comes into play.
Also, the assumption that the MTP structures are responsible for
particle production (this corresponds to the conventional 
linear approximation of the Landau-Zener model)
 makes our arguments very simple. 
What we have shown is that just one MTP of the turning point (3.)
 appearing among other turning points can generate
 matter-antimatter asymmetry. 

We have shown that interference between different Stokes phenomena is
essential for the asymmetry. 
In order for interference to occur, there must be a mixture of Stokes
phenomena with $\varphi_1$ solutions and $\varphi_2$ solutions. 
This criterion is the main result of this paper.
As far as we know, this is the first paper to point out the importance
of the interference between different Stokes phenomena
for the asymmetry of dynamical particle production.

Although the model examined in this paper is very simple, in which 
only a complex scalar field and the simplest interaction are considered,
we believe that the model has the essential properties of
dynamical particle production with CP-violating interactions and is
 giving the first example of the analysis in which the interference
 between different Stokes phenomena is taken into account for the
 higher-order differential equation.

\section{Conclusions and discussions}
In this paper, we have analyzed the mechanism of matter-antimatter
asymmetry generation during preheating when CP is broken.
We found that interference between different Stokes phenomena is
an important factor in asymmetry generation.
This criterion is the main result of this paper.
To the best of our knowledge, this is the first paper to point out 
that interference between different Stokes phenomena is 
crucial for the asymmetry in dynamical particle production.
Although the model considered in this paper is very simple, 
we believe it has the essential properties of
dynamical particle production with CP-violating interactions.

In our analysis, the quantization given by Eq.(\ref{eq-q-comp})
does not hold for the transition process.
Therefore, we discussed particle generation using the original
creation and annihilation operators of the real fields $\varphi_1$ and
$\varphi_2$.
What makes the analysis difficult is the fourth-order differential equation
that appears after decoupling the simultaneous differential equations.
This equation is so complex that we had to carefully pick out the essential
elements that describe the physics.
We followed the analysis of the multi-level Landau-Zener model.

The EWKB analysis considers the Borel resummation, and its Stokes lines
describe the global structure (i.e, the connection formulae).
Higher-order differential equations require new
Stokes lines and virtual turning points for global consistency.
The key questions for physics are (1) what Stokes phenomena are
responsible for (asymmetric) particle production, and (2)
whether the ``new'' objects in the higher-order differential equation are 
changing the traditional particle production scenario.
We then identified the Stokes phenomena responsible for
the asymmetry.
The matter-antimatter asymmetry requires mixing
between the asymptotic solutions of $\varphi_1$ and $\varphi_2$, which
is possible only when $\Lambda_I$ exists.
For our calculations, we had to draw a structure of the Stokes lines 
in order to understand which Stokes lines are responsible for the Bogoliubov
transformation of the model.
Unfortunately, at least in the simple model considered in this paper, 
all new Stokes lines are drawn as dotted lines on the real-time axis and 
play no significant role in physics.

We have used the idea that the situation near particle production 
is similar to the Landau-Zener transition;
the MTP gives a simple connection matrix, which is a simple and
straightforward way to calculate the local system. 
The basic calculations for the local system are similar to the conventional
preheating scenario or the Landau-Zener transition.
 
As a by-product, we have proven that a rotational motion, which is given by
$\Lambda(t)=\Lambda_0e^{i\theta(t)}$
or $\Lambda(t)=\Lambda_0e^{iA \cos(\omega t)}$, cannot generate
 matter-antimatter asymmetry in this model.
This tells us that expanding the interaction as
$\Lambda_0e^{iA \cos(\omega t)}\simeq \Lambda_0[1+iA\cos(\omega t)]$
for small $A$ is very dangerous.
In light of the EWKB, such a perturbative expansion can easily change the global
structure of the Stokes lines, generating false turning points on the
complex $t$-plane.

Finally, we will compare our present result with our earlier calculation
of asymmetric preheating\cite{Enomoto:2021hfv}.
In our previous calculation, we had considered time-dependent
off-diagonal elements of the simultaneous differential equation of a
Majorana fermion to find that the helicity asymmetry occurs
when particle production is not simultaneous between different
helicity states.
After decoupling, the equation of a Majorana fermion is the second-order
differential equation and there is no interference of the Stokes phenomena
between different species. 
In this paper, we have shown that the interference of the Stokes
phenomenon is the 
essence of asymmetry for a certain model, but we do not rule out other
possibilities for the asymmetry generation.

\acknowledgments
The authors would like to thank Nobuhiro Maekawa for his collaboration in
the very early stages of this work.
SE  was supported by the Sun Yat-sen University Science Foundation.
\appendix
\section{The Stokes phenomena for the Landau-Zener transition}
\label{app-zener}
We show how the Landau-Zener model\cite{Zener:1932ws}
can be related to the Stokes phenomena of cosmological particle
production.
The point is that the Stokes lines of the Landau-Zener transition give
the MTP structure at the transition, where the ``level crossing''
occurs.
For the extended Landau-Zener model with multiple levels, the MTP
structure appears at each level crossing.

We introduce the ``velocity'' $v>0$ and the off-diagonal elements
$\Delta$, both of which are supposed to be real.
The Landau-Zener model uses a couple of ordinary differential equations
given by
\begin{eqnarray}
i\hbar\frac{d}{dt}\left(
\begin{array}{c}
\psi_1\\
\psi_2
\end{array}
\right)&=&\left(
\begin{array}{cc}
-\frac{v}{2}t& \Delta \\
 \Delta& +\frac{v}{2}t 
\end{array}
\right)
\left(
\begin{array}{c}
\psi_1\\
\psi_2
\end{array}
\right),
\end{eqnarray}
which can be decoupled to give 
\begin{eqnarray}
\left[\hbar^2\frac{d^2}{dt^2}+\left(\Delta^2-i\hbar\frac{v}{2}\right)+\frac{1}{4}v^2t^2\right]\psi_1&=&0\\
\left[\hbar^2\frac{d^2}{dt^2}+\left(\Delta^2+i\hbar\frac{v}{2}\right)+\frac{1}{4}v^2t^2\right]\psi_2&=&0.
\end{eqnarray}
Following Refs.\cite{Virtual:2015HKT, Voros:1983, Delabaere:1993, Silverstone:2008},
 we are going to rewrite the
 equations in the standard EWKB form.
We have
\begin{eqnarray}
\left[-\frac{d^2}{dx^2}+\eta^2 Q(x)
\right]\psi(x,\eta)&=&0,
\end{eqnarray}
where
\begin{eqnarray}
Q(x)&\equiv&V(x)-E
\end{eqnarray}
is given by the ``potential'' $V$ and the ``energy'' $E$.
For the decoupled equations of the Landau-Zener model, we have
\begin{eqnarray}
Q(x,\eta)&=&\left(\Delta^2-i\eta^{-1}\frac{v}{2}\right)+\frac{1}{4}v^2t^2\nonumber\\
&=&\left(\Delta^2+\frac{1}{4}v^2t^2\right)+\left(\mp i\eta^{-1}\frac{v}{2}\right)\\
Q_0(x)&\equiv&\Delta^2+\frac{1}{4}v^2t^2\\
Q_{-1}(x)&\equiv&\mp i\eta^{-1}\frac{v}{2}.
\end{eqnarray}
Due to the formal structure of the EWKB, the exact Stokes lines are drawn
using only $Q_0$.
Therefore, using the EWKB formulation, one will find that $\psi_1$ and $\psi_2$ 
have the same Stokes lines.
(A careful reader will understand that this statement does not mean that
solutions are identical.) 
Finally, we have
\begin{eqnarray}
V&=&-\frac{1}{4}v^2x^2\\
E&=&\Delta^2 
\end{eqnarray}
for the conventional quantum scattering problem with an inverted
quadratic potential, which gives the MTP structure at the transition.
Note also that for $\Delta=0$ the MTP structure shrinks to be a double
turning point, from which four Stokes lines are coming out.

If one wants to consider (explicitly) the exact solution instead of the
Stokes lines of the EWKB, it will be convenient to consider 
$z=i\sqrt{v} e^{i\pi/4}t$
($z^2=-ivt^2$) to find\footnote{Here we temporarily set $\hbar=1$
because we are calculating the exact solution and considering no
expansion with respect to $\hbar$.}
\begin{eqnarray}
\left[\frac{d^2}{dt^2}+\left(n+\frac{1}{2}-\frac{1}{4}z^2\right)\right]\psi_1(z)&=&0\\
\left[\frac{d^2}{dt^2}+\left(n-\frac{1}{2}-\frac{1}{4}z^2\right)\right]\psi_2(z)&=&0.
\end{eqnarray}
Here we set 
\begin{eqnarray}
n&\equiv&i\frac{\Delta^2}{v}.
\end{eqnarray}
Since these equations are giving the standard form of the Weber
equation, their solutions are given by a couple of independent
combinations of $D_n(z), D_n(-z),D_{-n-1}(iz),
D_{-n-1}(-iz)$.
Using the asymptotic forms of the Weber function, one can easily get the
transfer matrix given by
\begin{eqnarray}
\left(
\begin{array}{c}
\psi_1^+\\
\psi_2^+
\end{array}
\right)&=&\left(
\begin{array}{cc}
e^{-\pi \kappa}& -\sqrt{1-e^{-2\pi\kappa}} \\
\sqrt{1-e^{-2\pi\kappa}} & e^{-\pi \kappa}
\end{array}
\right)
\left(
\begin{array}{c}
\psi_1^-\\
\psi_2^-
\end{array}
\right),\nonumber\\
\end{eqnarray}
where phase parameters are disregarded for simplicity.
$\pm$ signs of $\psi^\pm$ are for $t\rightarrow \pm \infty$.
We introduced $\kappa$, which is the imaginary part of $n$ and given by
\begin{eqnarray}
\kappa&\equiv&\frac{\Delta^2}{v}.
\end{eqnarray}

One might think that the result is trivially identical to the scattering
by the inverted quadratic potential, but it isn't.
Note that the above transfer matrix is not defined for the ``adiabatic
states'', which represent the ``adiabatic energy'' 
\begin{eqnarray}
E_\pm&=&\pm\sqrt{\Delta^2+v^2t^2/4}.
\end{eqnarray}
Since these adiabatic states are diagonalizing the Hamiltonian and 
identified with the asymptotic WKB
solutions, the transition matrix for these (adiabatic) states is giving
Bogoliubov transformation of the cosmological particle production.
If one writes the transfer matrix for these ``adiabatic states''
$\Psi_{1,2}$ instead of the original states $\psi_{1,2}$, 
one will have
\begin{eqnarray}
\left(
\begin{array}{c}
\Psi_1^+\\
\Psi_2^+
\end{array}
\right)&=&\left(
\begin{array}{cc}
 \sqrt{1-e^{-2\pi\kappa}} &e^{-\pi \kappa}\\
e^{-\pi \kappa} &-\sqrt{1-e^{-2\pi\kappa}}
\end{array}
\right)
\left(
\begin{array}{c}
\Psi_1^-\\
\Psi_2^-
\end{array}
\right),\nonumber\\
\end{eqnarray}
where we have recovered the result of the inverted
quadratic potential.
Here we have omitted the phase parameter.

One can easily compare the above transfer matrix with the bosonic
preheating of Ref.\cite{Kofman:1997yn}.
For Dirac fermions, one can find the calculation based on the 
Landau-Zener model in Ref.\cite{Enomoto:2020xlf},
 which can be compared with the standard
calculation of Refs.\cite{Greene:1998nh, Peloso:2000hy}.
For Majorana fermions, one can find the calculation in
Ref.\cite{Enomoto:2021hfv}, in which an asymmetry is also discussed in
detail.

The off-diagonal elements of the transfer matrix are giving
$\beta_k^+$ of the Bogoliubov transformation\cite{Kofman:1997yn}
if $\alpha_k^-=1, \beta_k^-=0$ is considered for the initial condition.

Comparing the original equation of the Landau-Zener model and
 the decoupled equations,
one can see that $D_1\equiv - vt, D_2\equiv +vt$ in the (original) 
diagonal elements are transferred into the
 ``potential'' $-\frac{1}{4}v^2t^2$ in the decoupled
 equations\cite{Enomoto:2020xlf}.
The Stokes lines are forming the MTP structure.
This explains the ``linear approximation'' used in this paper and 
in the original Landau-Zener model.

\end{document}